\documentclass[12pt]{article}
\pdfoutput=1
\usepackage{jheppub}

\usepackage{graphicx}

\usepackage{amsmath}

\usepackage{amssymb}
\usepackage{mathtools}

\usepackage{float}
\usepackage{empheq}

\newcommand{\be}{\begin{equation}}
\newcommand{\ee}{\end{equation}}
\newcommand{\ba}{\begin{aligned}}
\newcommand{\ea}{\end{aligned}}

\newcommand{\bz}{\bar{z}}

\usepackage{tensor}
\newcommand{\ind}[1]{\indices{#1}}

\title{A One-Loop Test of the near-AdS$_2$/near-CFT$_1$ Correspondence}
\author{Anthony M. Charles$^{a}$ and}
\author{Finn Larsen$^b$}

\affiliation{$^a$Institute for Theoretical Physics, KU Leuven, \\
Celestijnenlaan 200D, B-3001 Leuven, Belgium}

\affiliation{$^b$Department of Physics and Leinweber Center for Theoretical Physics, \\
University of Michigan, 450 Church Street, Ann Arbor, MI 48109-1020, USA}

\emailAdd{anthony.charles@kuleuven.be}
\emailAdd{larsenf@umich.edu}

\abstract{ We analyze quantum fluctuations around black hole solutions to the Jackiw-Teitelboim model. 
We use harmonic analysis on Euclidean AdS$_2$ to show that the logarithmic corrections to the partition function are determined 
entirely by quadratic holomorphic differentials, even when conformal symmetry is broken and harmonic modes are no longer true zero modes. Our quantum-corrected partition function agrees precisely with the SYK result. We argue that our effective quantum field theory methods and results generalize to other theories of two-dimensional dilaton gravity.
}

\begin{document}

\maketitle 

\section{Introduction and Summary}
\label{sec:intro}

The AdS$_{d+1}$/CFT$_d$ correspondence has proven to be an indispensable tool for understanding quantum gravity.  In particular, it is a strong/weak duality that gives unprecedented insights into the microscopic dynamics of black holes.  However, the special case of AdS$_2$/CFT$_1$ is 
poorly understood, for a variety of reasons.  The usual holographic decoupling limits fail when $d = 1$, global AdS$_2$ contains two disconnected boundaries (unlike its higher-dimensional counterparts), and conformal symmetry precludes any finite-energy excitations of a theory living on AdS$_2$~\cite{Maldacena:1998uz,Strominger:1998yg}.  Resolution of these issues would have fundamental implications for near-extremal black holes, as their near-horizon geometry almost universally contains an AdS$_2$ factor.

Recently, progress has been made by developing a near-AdS$_2$/near-CFT$_1$ duality, wherein a two-dimensional gravitational theory living on a background that approximates AdS$_2$ is dual to a one-dimensional quantum theory that is nearly conformal.  A concrete realization of this proposal is the Sachdev-Ye-Kitaev (SYK) model~\cite{Sachdev:1992fk,Kitaev}, a one-dimensional theory of Majorana fermions that has an emergent conformal symmetry in the infrared.  This conformal symmetry is both explicitly and spontaneously broken at finite temperature, and the goldstone bosons of the broken conformal symmetry are described by the Schwarzian action~\cite{Maldacena:2016hyu}.  At low energies, the SYK model is thought to be dual to a two-dimensional dilaton gravity theory that exhibits the same pattern of conformal symmetry breaking~\cite{Maldacena:2016upp,Jensen:2016pah}.  The Jackiw-Teitelbom (JT) model~\cite{Teitelboim:1983ux,Jackiw:1984je} represents 
a particularly simple universality class that is described by near-AdS$_2$/near-CFT$_1$ holography in this manner. Modern analyses of this model include~\cite{Maldacena:2016upp,Almheiri:2014cka,Almheiri:2016fws,Engelsoy:2016xyb,Grumiller:2017qao,Harlow:2018tqv,Castro:2018ffi}. 

The linchpin of the duality between the SYK and JT models is the classical equivalence of both theories to the Schwarzian.  If these theories are truly dual, however, they must be equivalent even at the quantum level. In particular, their one-loop partition functions must agree. Computations in the SYK model~\cite{Maldacena:2016hyu} found
\begin{equation}
	\log Z|_\text{one-loop} = - \frac{3}{2} \log \beta J~,
\label{eq:syk_z}
\end{equation}
where $\beta$ is the inverse temperature and $J$ the parameter in the SYK model that controls the Gaussian distribution of random couplings.  This same result can also be derived from the Schwarzian theory. We review the SYK and Schwarzian derivations in appendix~\ref{app:z}.  
Our goal in this paper is to explicitly quantize the JT model using Euclidean quantum gravity methods, compute its bulk one-loop partition function, and provide a precision test of the near-AdS$_2$/near-CFT$_1$ duality by demonstrating that it matches (\ref{eq:syk_z}).  

One approach to quantize the JT model is to classically recast the theory as a one-dimensional Schwarzian theory and then 
use the relative simplicity of this boundary theory to compute the quantum path integral exactly~\cite{Stanford:2017thb,Kitaev:2018wpr,Yang:2018gdb,Saad:2019lba,Iliesiu:2019xuh,Stanford:2019vob}.  These methods all yield a density of states for the JT model that is consistent with the one-loop SYK partition function (\ref{eq:syk_z}).  However, these approaches seem to rely heavily on the specifics of the JT model, and in particular on the dilaton appearing as a Lagrange multiplier that forces the background to have constant negative curvature. Compactification of gravitational theories down to two dimensions generically spoils these nice features through the introduction of non-trivial interactions between the dilaton and the other fields in the theory.  This motivates quantizing the JT model directly in the bulk without resorting to the Schwarzian boundary theory. The methodology we develop to do so applies also to more realistic theories, and in particular ones with direct realizations in string theory.

In our bulk analysis we consider explicit black hole solutions to the JT model and study quadratic fluctuations of the action around these backgrounds. This yields a theory of metric and dilaton fluctuations coupled by non-minimal (and fairly complicated) interactions that depend on the background dilaton. Moreover, the dilaton profile blows up at the AdS$_2$ boundary so we must carefully keep track of divergences when integrating over the entire spacetime volume. Without a road map to guide us on how to handle these challenges, explicit computation of the bulk one-loop partition function is a daunting task.  

Therefore, in section~\ref{sec:quant}, we analyze a simpler free model consisting of independent metric, dilaton, and ghost fields propagating on a non-dynamical Euclidean AdS$_2$ background, i.e. the two-dimensional hyperbolic plane $H^2$. In this simplified setting, vector and tensor fields are precisely equivalent to scalars, up to the addition of certain discrete modes that have no scalar analogue. This way of organizing the field content not only makes computing the required functional determinants straightforward but also makes manifest that these one-loop determinants cancel precisely against each other. This leaves only contributions from the discrete modes that completely determine the one-loop partition function in the simplified model.
 

Having addressed some of the central challenges in the context of a simplified model, in section~\ref{sec:oneloop} we adapt the reasoning to the the full JT model. Even with the addition of non-minimal couplings that depend on the background dilaton, it is useful to represent all fields as scalars, up to the discrete modes. The dilaton profile, which we interpret as a thermal background due to the nearby black hole, obstructs the precise cancellations we established in the simplified model, but only up to terms that are quadratic in temperature. Therefore the quantum correction will once again be dominated by the discrete modes and their contributions sum up to a logarithmic term in the one-loop partition function for the full JT model that precisely matches the SYK result (\ref{eq:syk_z}).

Several distinct types of discrete modes enter our computations but the central ones are the quadratic holomorphic differentials. These discrete modes are deformations of the metric that can be formally represented as pure diffeomorphisms, but they are nonetheless physical because the required diffeomorphisms are non-normalizable. The
quadratic holomorphic differentials can exist only because AdS$_2$ is noncompact, so they have no analogues on the conformal disc. Instead, they are close relatives of the Brown-Henneaux deformations that yield the central charge in the AdS$_3$/CFT$_2$ correspondence. Moreover, they are dual to the soft reparametrization modes described by the Schwarzian action, so they offer a satisfying connection with other approaches.

Our computation is similar in spirit to previous evaluations of logarithmic corrections to extremal black hole entropy~\cite{Banerjee:2010qc,Banerjee:2011jp,Sen:2011ba,Keeler:2014bra,Larsen:2014bqa}. 
Such logarithmic terms arise entirely from quantum fluctuations of fields in the near-horizon AdS$_2$ geometry of the black holes. Presently, the background is only ``nearly" AdS$_2$, and so conformal symmetry is broken. Our result is that, despite this difficulty, we retain sufficient control to reliably compute the logarithmic terms in the one-loop partition function and reproduce (\ref{eq:syk_z}).  

The organization of this paper is as follows.  In section~\ref{sec:jt}, we review semi-classical black holes in the JT model and their spontaneously broken conformal symmetry.  In section~\ref{sec:quant}, we study quantization in AdS$_2$ and explicitly detail how to 
dualize the metric and other quantum fields to scalars, up to a particular discrete set of modes that have no scalar analogue.  
We contrast quantization in AdS$_2$ with standard worldsheet methods in string theory and identify discrete modes with quadratic holomorphic differentials that have no analogues on compact Riemann surfaces. 
In section~\ref{sec:oneloop}, we carefully apply these methods to computing the one-loop partition function of black holes in the JT model.  In particular, we show that the one-loop correction is entirely due to the discrete modes, the aforementioned quadratic holomorphic differentials, and it precisely matches the SYK result (\ref{eq:syk_z}).  In section~\ref{sec:bhentropy}, we show that the corresponding one-loop quantum corrections to black hole entropy vanish and we discuss the implications.  Finally, we conclude in section~\ref{sec:conc} by discussing how our methods contrast with the Schwarzian computation and generalize to other theories of nearly-AdS$_2$ gravity.

\section{The Jackiw-Teitelboim Model}
\label{sec:jt}

In this section we review black hole solutions in the Jackiw-Teitelboim model and define a precise limit in which they are described semi-classically by effective field theory. 
We then study Killing vectors on these black hole backgrounds and show how they fit into the conformal isometry group that is spontaneously broken at finite temperature. 

\subsection{Action and Equations of Motion}

The Jackiw-Teitelboim model is the dilaton gravity theory in $1+1$ dimensions with action
\begin{equation}
	I_\text{JT} = -\frac{1}{16\pi G_2}\int d^2x\,\sqrt{g}\left(\Phi R - \frac{2}{\ell^2} + \frac{2 \Phi}{\ell^2} \right) - \frac{1}{8\pi G_2}\int dt\,\sqrt{\gamma}\,\Phi K~,
\label{eq:sjt}
\end{equation}
where $G_2$ is Newton's gravitational coupling constant in two dimensions, $\Phi$ is the dilaton, and $\ell$ is a length scale that will shortly be identified with the AdS$_2$ radius. In the Gibbons-Hawking-York boundary term $\gamma_{ab}$ is the induced metric on the boundary and $K$ is the extrinsic curvature.  This term imposes boundary conditions on the bulk fields and makes the variational principle well-posed.  

The classical equations of motion are
\begin{equation}
	R = -\frac{2}{\ell^2}~, \quad \nabla_\mu \nabla_\nu \Phi = \frac{g_{\mu\nu}}{\ell^2}\left(\Phi - 1 \right)~,
	\label{eqn:bkndeom}
\end{equation}
and so the background has constant negative curvature with radius $\ell$. The Jackiw-Teitelboim model realizes the 
symmetry breaking pattern of the original SYK model, where the zero modes in the infrared are Nambu-Goldstone modes of the broken conformal symmetry and described by an effective Schwarzian action~\cite{Maldacena:2016upp}.

We interpret the Jackiw-Teitelboim model as a low-energy effective theory of gravity that arises by compactification of a higher-dimensional one.  Then the scale $\Lambda_\text{KK}$ of the internal manifold is related to the  length scale $\ell\sim \Lambda^{-1}_\text{KK}$ of the background, i.e. the massive Kaluza-Klein modes have Compton wavelengths of the order of the AdS$_2$ length scale. 

In generic compactifications there will be more matter fields. If the additional matter is minimally coupled to the gravitational sector, the dilaton equation of motion becomes
\begin{equation}
	\nabla_\mu \nabla_\nu \Phi - \frac{g_{\mu\nu}}{\ell^2}\left(\Phi - 1\right) = - 8 \pi G_2 T_{\mu\nu}~,
\label{eq:dileom_t}
\end{equation}
where $T_{\mu\nu}$ is the energy-momentum tensor of the additional matter.  In this and more elaborate settings the effective 
Schwarzian theory~\cite{Liu:2019niv,Sachdev:2019bjn} and its SYK-like dual~\cite{Mertens:2019tcm} must both be modified. The resulting logarithmic quantum correction \eqref{eq:syk_z} generally changes as well. We restrict ourselves to the ``pure" Jackiw-Teitelboim model with no additional matter fields, but we foresee no difficulties in generalizing our computations to other situations.

\subsection{Black Hole Solutions}
\label{sec:jt:vacuum}

A metric on a two-dimensional manifold with Lorentzian signature and constant, negative curvature $R = -\frac{2}{\ell^2}$ can be presented in Poincar\'e coordinates as
\begin{equation}
	ds^2 = \frac{\ell^2}{x^2}\left(-dt^2 + dx^2\right) = -\frac{4 \ell^2 dx^+ dx^-}{(x^+ - x^-)^2}~,
\end{equation}
with light-cone coordinates $x^\pm = t \pm x$. We follow the conventions of~\cite{Almheiri:2014cka} and define the coordinates $x$, $t$ to be dimensionless, in order to keep the AdS$_2$ length scale $\ell$ explicit in all expressions. The spatial coordinate $x$ lies in the range $x \in (0, \infty)$, with $x = 0$ corresponding to the boundary of the manifold.  In the absence of matter, the classical dilaton equation of motion on this background \eqref{eq:dileom_t} can be integrated exactly with the result
\begin{equation}
	\Phi = 1 + \frac{a - \mu x^+ x^-}{x^+ - x^-}~,
\label{eq:dil_class}
\end{equation}
where $a$, $\mu$ are dimensionless constants.  The parameter $a$ is interpreted as the gravitational backreaction even though, in the JT-model, the AdS$_2$ geometry is exact. The background nearly preserves AdS$_2$ symmetries in the region $x \gg a$, where the dilaton profile remains small $|\Phi-1|\ll 1$ .  When $x \ll a$, though, the dilaton profile blows up and then the classical solution entirely invalidates the AdS$_2$ symmetry.  

We require that $a > 0$ and $\mu \geq 0$ and interpret the solution as a black hole with a horizon along the null line $x^+ = \sqrt{a/\mu}$~\cite{Almheiri:2014cka}.  Moreover, for $a > \mu$, the singularity at the center of the black hole is time-like.  The mass $M$ (above extremality) and temperature $T$ of the black hole are given by
\begin{equation}
	M = \frac{\mu}{8 \pi G_2 \ell}~, \quad T = \frac{1}{\pi \ell}\sqrt{\frac{\mu}{a}}~.
\label{eq:bhmass}
\end{equation}
For black holes in general theories of two-dimensional dilaton gravity, the black hole entropy is given by~\cite{Grumiller:2007ju}
\begin{equation}
	S = \frac{\Phi|_H}{4 G_2}~,
\label{eq:bhlaw}
\end{equation}
where $\Phi|_H$ is the value of the dilaton at the horizon of the black hole.  This quantity plays the role of the horizon area for these two-dimensional black holes.  In particular, for black holes in the Jackiw-Teitelboim model, the entropy is
\begin{equation}
	S = S_0 + \Delta S = \frac{1}{4G_2} + \frac{\sqrt{\mu a}}{4G_2}~,
\end{equation}
where we have explicitly separated out the extremal entropy $S_0$ and the additional contribution $\Delta S$ at finite temperature.

\subsection{The Semi-Classical, Near-Extremal Regime}
\label{sec:jt:scales}

Extremal black holes have vanishing temperature and in the near-extreme regime 
\begin{equation}
	\Delta S = S - S_0 \ll S_0  ~,
	\label{eqn:nearextS}
\end{equation}
we can expand the entropy and present it as a linear function of the small temperature. 
The first law of thermodynamics then relates the mass and entropy above extremality so that they take the form
\begin{equation}
	M  = \frac{T^2}{2M_\text{gap}}~, \quad S - S_0 = \frac{T}{M_\text{gap}}~,
\label{eq:above_ext}
\end{equation}
for some energy scale $M_\text{gap}$.\footnote{Equivalently, we could also quantify how close we are to extremality through the length scale $L =  \frac{2}{\pi} \frac{\partial S}{\partial T} = \frac{2}{\pi}M_\text{gap}^{-1}$.  The normalization choice of $\frac{2}{\pi}$ guarantees that this length scale coincides with the long string scale~\cite{Larsen:2018iou}. \cite{Sachdev:2019bjn} uses the notation $\gamma=M^{-1}_{\rm gap}$ and \cite{Hong:2019tsx} has $L=M^{-1}_{\rm gap}$. }   For the class of black holes in two dimensions analyzed in this paper the mass gap is given by
\begin{equation}
	M_\text{gap} = \frac{4 G_2}{a \pi \ell}~.
\end{equation}

For black holes with mass below this scale $M \ll M_\text{gap}$, the energy of the black hole is \emph{smaller} than their temperature.  That is, below the mass gap, the black hole does not have sufficient energy to emit a single quantum of Hawking radiation with energy of order the temperature.  Therefore, the usual semi-classical understanding of black hole thermodynamics cannot apply in this regime \cite{Preskill:1991tb}.  In the pure AdS$_2$ limit where the dimensionless parameter $a \to 0$ the mass gap blows up and there is no regime of applicability for semi-classical physics.  A non-trivial dilaton profile is therefore mandatory. 

\begin{figure}[ht]\centering
	\includegraphics[scale = 0.7]{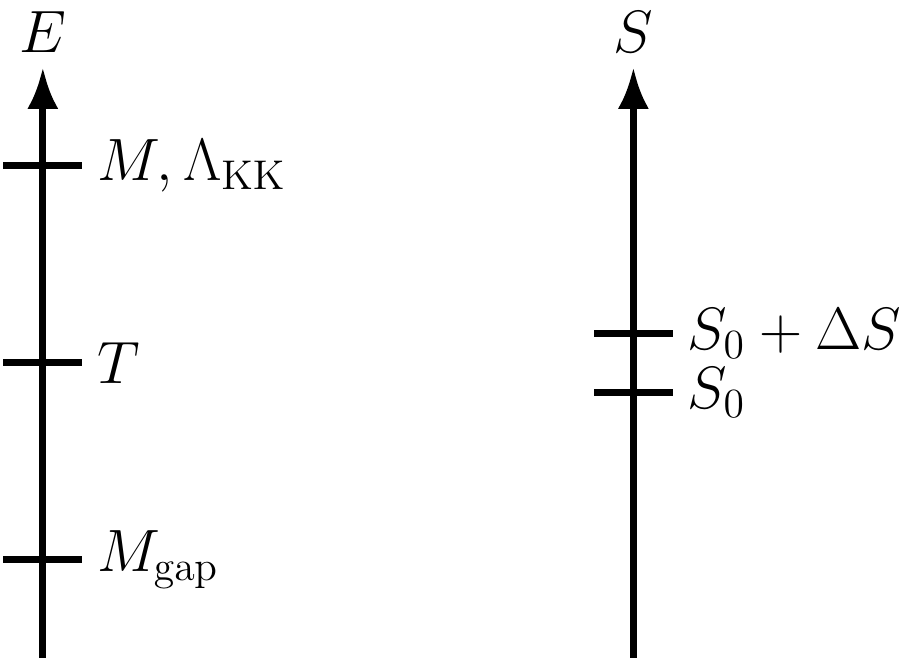}
	\caption{A schematic diagram of the hierarchy of different scales in our setup.  This choice in scales ensures that the Jackiw-Teitelboim model is an effective field theory that can describe semi-classical, near-extremal black holes.}
	\label{fig:scales}
\end{figure}

In addition to keeping the temperature well above the mass gap, the effective field theory of dilaton gravity is valid only in the regime where thermal fluctuations are well below the compactification scale $\Lambda_\text{KK}$; otherwise, we would have to account for relatively large probabilities of exciting massive Kaluza-Klein modes.  We therefore require the hierarchy of scales 
\begin{equation}
	M_\text{gap} \ll T \ll \Lambda_\text{KK} \sim \frac{1}{\ell}~.
	\label{eqn:hier1}
\end{equation}
This hierarchy, when combined with the relationship (\ref{eq:above_ext}) for how mass scales with temperature, implies $T \ll M$. Recalling that by mass $M$ we refer to the energy above the extremal mass, we interpret our background as a large, near-extremal black hole that slowly emits Hawking radiation. Indeed, this is the regime that is relevant for precision studies of near-extremal black hole thermodynamics~\cite{Castro:2009jf,Pathak:2016vfc,Castro:2018ffi}.

Our hierarchies are summarized in figure~\ref{fig:scales}. They can be concretely realized in our model by choosing the parameters $a$ and $\mu$ such that
\begin{equation}
	\mu \ll a\,\,\,\,\text{and}\,\,\,\, G_2 \ll \sqrt{\mu a} \ll 1~.
	\label{eqn:hier2}
\end{equation}
It is the requirement \eqref{eqn:nearextS} that the ground-state entropy of the black hole is large compared to the correction arising from finite-temperature effects 
that demands the parameter $\sqrt{\mu a}$ to be small. This condition enforces that the geometry really is nearly AdS$_2$  and offers a quantitative measure that conformal symmetry is only mildly broken. We will later use the smallness of $\sqrt{\mu a}$ to justify taking finite-temperature effects into account 
perturbatively. 

In the extreme limit where the small parameter $\sqrt{\mu a}=0$, we have $\mu=0$, $a\neq 0$ and the background corresponds to the extremal black hole ground state. It is described by a dilaton profile that is non-trivial but static in Poincar\'e coordinates. This is a physical regime and violates the first inequality in the second equation of \eqref{eqn:hier2} only because the thermodynamic description is invalid in the extremal limit.

\subsection{Global Coordinates, Killing Vectors, and Conformal Symmetry}
\label{sec:jt:euclid}

Semi-classical computations in Euclidean quantum gravity are most convenient in global coordinates, where the black hole solution from subsection~\ref{sec:jt:vacuum} becomes~\cite{Almheiri:2016fws}
\begin{equation}
	ds^2 = \frac{4 \mu \ell^2}{a} \sinh\left(2\sqrt{\frac{\mu}{a}} x\right)^{-2}\left(-dt^2 + dx^2\right)~,~ \Phi = 1 + \sqrt{\mu a}\,\coth\left(2\sqrt{\frac{\mu}{a}} x\right)~.
\end{equation}
Euclideanization sends $t \to i \tau$, where $\tau$ is the (dimensionless) Euclidean time with period $\beta/\ell$ 
and $\beta=T^{-1}$ is the inverse of the temperature given in \eqref{eq:bhmass}. By further changing the variables to
\begin{equation}
	z = \text{exp}\left(-2\sqrt{\frac{\mu}{a}}x + 2i \sqrt{\frac{\mu}{a}} \tau\right)~,
\end{equation}
we can map the solution onto the disk in holomorphic coordinates:
\begin{equation}
	ds^2 = \frac{4\ell^2 dz d\bar{z}}{(1-z \bar{z})^2}~, \quad \Phi = 1 + \sqrt{\mu a}\left( \frac{1 + z \bar{z}}{1 -z \bar{z}}\right)~,
\label{eq:disk_metric}
\end{equation}
where the boundary is at $|z|=1$.  

We have picked our coordinate frame to bring the dilaton into the form given in (\ref{eq:disk_metric}), but there are actually three linearly independent dilaton profiles $X$, given by
\begin{equation}
	X_{-1} = \frac{2z}{1- z \bz}~, \quad X_{0} = \frac{1 + z \bz}{1- z \bz}~, \quad X_{+1} = \frac{2\bz}{1- z \bz}~,
\label{eq:xprofiles}
\end{equation}
such that $\Phi = 1 + X$ satisfies the background equations of motion \eqref{eqn:bkndeom}. These profiles reflect an underlying symmetry. They generate the three Killing vectors
\begin{equation}\begin{aligned}
	\zeta_n^\mu &= \epsilon^{\mu\nu} \nabla_\nu X_n = 
	\begin{dcases} 
		(z^2,-1) & n = -1~, \\
		(z,-\bar{z}) & n = 0~,\\
		(1,-{\bar z}^2) & n = +1~, 
	\end{dcases}
\label{eq:zetavectors}
\end{aligned}\end{equation}
corresponding to the $SL(2,\mathbb{R})$ isometry group of AdS$_2$. We can see this explicitly by defining the operators $L_n$ as Lie derivatives with respect to the Killing vector fields:
\begin{equation}
	L_n = \ell^2\mathcal{L}_{\zeta_n} = \ell^2(\epsilon^{\mu\nu}\nabla_\nu X_n)\nabla_\mu~.
\end{equation}
These operators satisfy the global $SL(2,\mathbb{R})$ algebra
\begin{equation}
	[L_n, L_m] = (n-m) L_{n+m}~.
	\label{eqn:Killingalg}
\end{equation}
The three vector fields $\zeta_n^\mu$ with $n=-1, 0, 1$ are the only Killing vectors on AdS$_2$. 

The three dilaton profiles \eqref{eq:xprofiles} not only yield the Killing vectors \eqref{eq:zetavectors} on AdS$_2$, but the 
equation of motion \eqref{eqn:bkndeom} for the background dilaton also ensures that the 
vector field $\nabla^\mu\Phi$ satisfies the {\it conformal} Killing vector equation. Thus the vector fields
\begin{equation}\begin{aligned}
	\xi_n^\mu &= \nabla^\mu X_n =\begin{dcases} 
		(z^2,1) & n = -1~, \\
		(z,\bar{z}) & n = 0~,\\
		(1,{\bar z}^2) & n = +1~, 
	\end{dcases}
\label{eq:xivectors}
\end{aligned}\end{equation}
are conformal Killing vectors. Moreover, they are just the first few entries in an infinite tower of conformal Killing vectors $\xi^\mu_n = (z^{-n+1},{\bar z}^{n+1})$ for any $n \in \mathbb{Z}$.  
This result is not special to AdS$_2$; any Riemannian manifold is conformally flat in two dimensions so there is an analogous tower of conformal Killing vectors on the background geometry in any two-dimensional theory of gravity.  

However, there is an important caveat. Conformal Killing vectors necessarily satisfy the conformal Killing vector equation, but this condition is not sufficient;
they must also be globally well-defined. For example, on the conformal disc the only true conformal Killing vectors among the entire tower of ``local" conformal Killing vectors 
are the three given in \eqref{eq:xivectors}. Moreover, in this aspect AdS$_2$ is {\it not} conformally equivalent to the disc. There are {\it no} normalizable conformal Killing vectors in AdS$_2$, due to its diverging conformal factor.

It is natural to extend the global $SL(2,\mathbb{R})$ algebra \eqref{eqn:Killingalg} formed from the Killing vectors by exploiting the conformal Killing vector fields.  
To do so, we define the operators $R_n$ as Lie derivatives with respect to the conformal Killing vectors with $n = -1, 0, +1$: 
\begin{equation}
	R_n = \ell^2\mathcal{L}_{\xi_n} = (\nabla^\mu X_n) \nabla_\mu~.
\end{equation}
These new conformal operators do not commute with the generators of the isometry algebra.  Instead, the algebra \eqref{eqn:Killingalg} gets extended as:
\begin{equation}\begin{aligned}
	[R_n, R_m] &= (n-m) L_{n+m}~, \\
	[L_n, R_m] &= (n-m) R_{n+m}~.
\label{eq:sl2sl2}
\end{aligned}\end{equation}
By defining the operators $J^\pm_n = \frac{1}{2}\left(L_n \pm  R_n\right)$ we see that the full conformal isometry algebra is two copies of the 
original $SL(2,\mathbb{R})$ algebra of isometries:
\begin{equation}
	[J^\pm_n, J^\pm_m] = (n-m) J^\pm_{n+m}~, \quad [J^+_n, J^-_m] = 0~.
\end{equation}
We define the quadratic Casimir $\mathbf{L}^2$ of the isometry algebra as 
\begin{equation}
\mathbf{L}^2 \equiv L_0^2 - \frac{1}{2}\left( L_{-1} L_{+1} + L_{+1} L_{-1}\right)~,
\end{equation}
and, by analogy, we introduce an operator that is quadratic in the conformal Killing vectors as
\begin{equation}
	\mathbf{R}^2 \equiv R_0^2 - \frac{1}{2}\left( R_{-1} R_{+1} + R_{+1} R_{-1}\right)~.
\end{equation}
Neither of these quadratic operators are Casimirs of the full $SL(2,\mathbb{R}) \times SL(2,\mathbb{R})$ algebra, but they are related to one another and to the Laplacian $\square \equiv \nabla_\mu \nabla^\mu$ by
\begin{equation}\begin{aligned}
	\ell^2 \square = \mathbf{L}^2 = -\mathbf{R}^2~.
\label{eq:l2r2}
\end{aligned}\end{equation}

This process of extending the isometry algebra into a conformal isometry algebra is well-known in the context of spin and quantum mechanics.  The $SU(2)$ algebra of spin raising and lowering operators can be extended to an $SU(2) \times SU(2) \cong SO(4)$ algebra by additionally quantizing components of the Laplace-Runge-Lenz vector~\cite{doi:10.2991/jnmp.2003.10.3.6}.  The $SL(2,\mathbb{R}) \times SL(2,\mathbb{R})$ algebra we find here can be obtained by analytic continuation of this $SO(4)$, as shown in appendix~\ref{app:cont}.

It was argued in~\cite{Strominger:1998yg} that for any quantum gravity theory on AdS$_2$, the classical global $SL(2,\mathbb{R})$ isometry group will be enhanced to the full Virasoro group.  This should be thought of as the two-dimensional analogue of the Brown-Henneaux mechanism that enhances 
global conformal symmetry $SO(2,2)$ to the full local conformal group ${\rm Vir}^2$ in AdS$_3$/CFT$_2$~\cite{Brown:1986nw}. In our discussion of AdS$_2$, the Killing vectors with $n=-1,0,1$ are enhanced to a full tower with any $n$. In the algebraic language developed in this subsection it is a diagonal subgroup
$SL(2,\mathbb{R})\subset SL(2,\mathbb{R}) \times SL(2,\mathbb{R})$ that is enhanced to Virasoro. 

At this point, we can explain precisely how bulk conformal symmetry is spontaneously broken in the Jackiw-Teitelboim model.  
For $\sqrt{\mu a} = 0$ the dilaton is constant $\Phi=1$ in global coordinates and the AdS$_2$ spacetime exhibits its full $SL(2,\mathbb{R})$ isometry. In contrast, at finite temperature  $\sqrt{\mu a} > 0$ the dilaton must be non-trivial and the profiles allowed by the equations of motion transform according to the spacetime $SL(2,\mathbb{R})$ isometry. Therefore, picking a particular profile for the dilaton is equivalent to picking a preferred direction within the global $SL(2,\mathbb{R})$ subgroup of the full conformal group.

The details relating the dilaton profiles to the full $SL(2,\mathbb{R}) \times SL(2,\mathbb{R})$ conformal isometry algebra developed in this subsection will play a central role in the remainder of the article. That is because the quantum fluctuations of the graviton and dilaton around the classical black hole background will naturally organize into representations of this algebra, even though it is broken.  In the low-temperature regime, the conformal symmetry breaking is mild enough that this classification will greatly simplify the accounting for finite-temperature effects in the one-loop partition function.

\section{Quantizing Fields in \texorpdfstring{AdS$_2$}{AdS2}}
\label{sec:quant}

In this section we study the quantum Jackiw-Teitelboim model without taking the dilaton profile into account. This amounts to a discussion of 
minimally-coupled fields propagating on Euclidean AdS$_2$, i.e. the two-dimensional hyperbolic space $H^2$.  We detail how vector and tensor fields are equivalent to scalar fields, up to the crucial addition of very specific discrete modes.  This way of organizing the field content makes many simplifications in the spectrum manifest and it becomes straightforward to 
compute functional determinants over the field fluctuations.

\subsection{Metrics and Normalizations}

In this section we will frequently use standard global coordinates where the $H^2$ metric is
\begin{equation}
	ds^2 = \ell^2\left(d \eta^2 + \sinh^2\eta\,d\phi^2\right)~.
\label{eq:dsh2}
\end{equation}
By making the change of variables $z = \tanh(\eta/2)e^{i\phi}$, we can also go to the conformal disk metric in holomorphic coordinates:
\begin{equation}
	ds^2 = \frac{4 \ell^2 dz d\bz}{\left(1- |z|^2\right)^2}~,
\end{equation}
where the boundary is at $|z| = 1$.  We will make use of both of these coordinate frames.  

We define the inner products for scalar fields $\phi$, vector fields $A_\mu$, and tensor fields $H_{\mu\nu}$ on the disk as:
\begin{equation}\begin{aligned}
\label{eqn:normalize}
	\langle \phi | \phi' \rangle &= \int d^2x\,\sqrt{g}\, \phi \phi'~, \\
	\langle A | A' \rangle &= \int d^2x\,\sqrt{g}\, g^{\mu\nu} A_\mu A_\nu'~, \\
	\langle H | H'\rangle &= \int d^2x \, \sqrt{g}\, g^{\mu\nu} g^{\rho \sigma} H_{\mu\rho} H_{\nu\sigma}'~.
\end{aligned}\end{equation}
It is significant that for fields with higher spin these normalization measures  include more factors of the inverse metric. Since the metric diverges at the boundary of AdS$_2$ such fields face weaker fall-off conditions near the boundary. 

\subsection{Scalar Fields in \texorpdfstring{AdS$_2$}{AdS2}}
\label{sec:quant:clas}

Consider a minimally coupled scalar field $\phi$ in AdS$_2$. It has kinetic operator $\Delta^{(0)} = -\square \equiv - \nabla_\mu \nabla^\mu$
with the eigenvalue equation
\begin{equation}
	\Delta^{(0)} \phi = \lambda^2 \phi~.
\label{eq:euclid_eig}
\end{equation}
In thermal AdS$_2$ with inverse temperature $\beta$, the normalizable eigenfunctions have discrete spectrum $\lambda^2 = \frac{n^2}{\beta^2} + \frac{1}{4\ell^2}$ for any integer $n$.  The spectrum on global AdS$_2$ (the covering space of AdS$_2$) is obtained by taking the zero-temperature limit where the eigenvalues become continuous and
\begin{equation}
	\lambda^2 = \frac{1}{\ell^2}\left(p^2 + \frac{1}{4}\right)~,
\end{equation}
for any real number $p$.  The eigenfunctions in global coordinates are given by~\cite{Camporesi:1994ga}
\begin{equation}\begin{aligned}
	u_{p m}(\eta, \phi) &= \frac{1}{\sqrt{2\pi}} \frac{1}{2^{|m|} |m|!} \bigg{|} \frac{\Gamma(ip + \frac{1}{2} + |m|)}{\Gamma(ip)}\bigg{|} e^{i m \phi} \sinh^{|m|}\eta \\
	&\quad \times{}_{2}F_1\left(ip + \frac{1}{2} + |m|, -i p + \frac{1}{2} + |m|, 1 + |m|, - \sinh^2\frac{\eta}{2}\right)~,
\label{eq:upm}
\end{aligned}\end{equation}
with $m$ taking integer values.  The normalization of these functions is chosen such that $\langle u_{p m} | u_{p' m'} \rangle = \ell^2 \delta(p - p') \delta_{m,m'}$.  The spectral density (also known as the Plancherel measure) of these modes is
\begin{equation}
	\mu(p) = \frac{p \tanh(\pi p)}{2\pi}~.
	\label{eqn:Planch}
\end{equation}
In the proceeding work, we will refer to these scalar configurations as \emph{continuous modes}, in order to emphasize the continuous nature of their spectrum.

The continuous modes comprise an irreducible representation of the $SL(2,\mathbb{R})$ isometry group of the background.  Following~\cite{Kitaev:2017hnr}, they form the principal unitary series and thus transform under $SL(2,\mathbb{R})$ generators (detailed in subsection~\ref{sec:jt:euclid}) as:
\begin{equation}\begin{aligned}
	\mathbf{L}^2 |u_{p m}\rangle &= -\left(p^2 + \frac{1}{4}\right)|u_{p m}\rangle~, \\
	L_0 |u_{p m}\rangle &= -m |u_{p m}\rangle~, \\
	L_{\pm 1} |u_{p m}\rangle &= - \left| ip \pm \frac{1}{2} + m \right| |u_{p m\pm 1}\rangle~.
\label{eq:rep}
\end{aligned}\end{equation}
The fact that the modes fall into such a representation is required purely by symmetry considerations, since the background $SL(2,\mathbb{R})$ symmetry is unbroken in this zero-temperature limit.

There are additional solutions to the eigenvalue equation (\ref{eq:euclid_eig}), but the corresponding eigenfunctions are not normalizable.  For example, any holomorphic (or anti-holomorphic) function will be a zero mode of the scalar Laplacian and thus satisfy (\ref{eq:euclid_eig}) with $\lambda^2 = 0$ because the scalar Laplacian on the conformal disk is given by
\begin{equation}
	-\square = -\frac{1}{\ell^2}\left(1 - |z|^2\right) \partial_z \partial_{\bz}~.
\end{equation}
The canonical modes $u_n = z^n$ (and their anti-holomorphic conjugates) form a complete set of zero modes but they are non-normalizable for any $n$
since $|u_n|^2 \to 1$ as $|z| \to 1$ and the AdS$_2$ volume diverges. 

Although holomorphic modes are not normalizable on Euclidean AdS$_2$ it is worth stressing that the modes $u_n = z^n$ {\it are} normalizable on the disc. This is possible even though these two geometries are conformally equivalent,
because the conformal transformation relating them is singular on the boundary and the normalization condition \eqref{eqn:normalize} is {\it not} conformally invariant for a scalar field. 

\subsection{Dualization of Vectors and Tensors to Scalars}
\label{sec:quant:dual}

The Euclidean eigenvalue equation for a vector field $A_\mu$ in AdS$_2$ is
\begin{equation}
	\Delta^{(1)} A_\mu \equiv \left(-\square - \frac{1}{\ell^2}\right) A_\mu = \lambda^2 A_\mu~.
\label{eq:eig_vector}
\end{equation}
It is significant that the Laplacian $\Delta^{(1)}$ differs from $-\square \equiv - \nabla_\mu \nabla^\mu$ by a term $-\frac{1}{\ell^2}$ due to the curvature of AdS$_2$.

In two Euclidean dimensions there is a canonical correspondence between the spectrum of the vector field and the spectrum of two scalars. 
However, the subtlety known as quantum inequivalence obstructs complete dualization of a vector into two scalars ~\cite{1980PhLB...94..179D,
Bastianelli:2005vk,Bastianelli:2005uy,Larsen:2015aia}. To explain, consider the Hodge decomposition stating that we can uniquely write the vector field as
\begin{equation}
	A_\mu = \nabla_\mu \phi_{\parallel} + \epsilon_{\mu\nu} \nabla^\nu \phi_\perp + \mathcal{A}_\mu~,
\end{equation}
where $\phi_{\parallel}$ and $\phi_{\perp}$ are scalar fields while $\mathcal{A}_\mu$ is a harmonic vector field.  Disregarding the latter momentarily, 
the eigenvalue equation \eqref{eq:eig_vector} becomes
\begin{equation}
	\nabla_\mu \left(-\square - \lambda^2 \right)\phi_{\parallel} + \epsilon_{\mu\nu}\nabla^\nu \left(-\square - \lambda^2 \right) \phi_{\perp} = 0~.
\end{equation}
Orthogonality then requires that both $\phi_{\parallel}$ and $\phi_\perp$ satisfy the scalar eigenvalue equation (\ref{eq:euclid_eig}) with the same value of $\lambda^2$
as the vector field $A_\mu$. 

If this was the complete story, then a vector field in two dimensions would be exactly equivalent to two scalar fields. However, the 
harmonic mode $\mathcal{A}_\mu$ requires special consideration. By definition, it is a zero mode of the vector field kinetic operator
\begin{equation}
	\Delta^{(1)}\mathcal{A}_\mu = 0~.
\end{equation}
The harmonic vector is a field configuration that can be dualized to \emph{either} of the two scalar modes so we should be careful to not over-count such modes. More importantly, the required dual scalar must satisfy the harmonic condition $\square \phi = 0$. We discussed the candidate harmonic scalars in the end of the preceding subsection and stressed that 
such scalar zero modes are non-normalizable, and so they are not truly zero modes. However, the vector field $\mathcal{A}_\mu = \nabla_\mu \phi_{\parallel}$ (or $\mathcal{A}_\mu = \epsilon_{\mu\nu}\nabla^\nu\phi_{\perp}$) \emph{is} normalizable.  Thus the spectrum of the vector field on $H^2$ is equivalent to two scalar fields and \emph{in addition} includes physical zero modes with no analogue in the scalar spectrum. Explicitly, the holomorphic scalars
\begin{equation}
	u_n = \frac{1}{\sqrt{2\pi n}}z^n~,
\end{equation}
with $n = 1,2,\ldots$ generate properly normalized harmonic modes for the vector field:
\begin{equation}
	\mathcal{A}_z^{(n)} = \ell\nabla_z u_n~, \quad \langle \mathcal{A}^{(m)} | \mathcal{A}^{(n)} \rangle= \ell^2 \delta_{m,n}~.
\label{eq:harmonic_basis}
\end{equation}

We can analyze a symmetric traceless tensor $H_{\mu\nu}$ on AdS$_2$ similarly. It has eigenvalue equation
\begin{equation}
	\Delta^{(2)} H_{\mu\nu} \equiv \left(-\square - \frac{4}{\ell^2}\right) H_{\mu\nu} = \lambda^2 H_{\mu\nu}~.
\label{eq:eig_tensor}
\end{equation}
It is again significant that the Laplacian $\Delta^{(2)}$ differs from $-\square \equiv - \nabla_\mu \nabla^\mu$ by a term $-\frac{4}{\ell^2}$ due to the curvature AdS$_2$.

In this case there is a correspondence between the spectrum for the symmetric traceless tensor and the spectrum of a vector, including a simple map between the corresponding eigenfunctions.  This correspondence is one-to-one on all non-harmonic modes as well as harmonic modes with $\lambda^2 = 0$ that appear in both cases.  However, the tensor also has ``higher" harmonic modes with eigenvalue $\lambda^2 = - \frac{2}{\ell^2}$ that have no analogue in the vector spectrum.

The idea that establishes these claims is that essentially all symmetric traceless tensors in two dimensions can be presented formally as pure diffeomorphisms.  Accordingly, consider a vector $\xi_\mu$ that solves the vector eigenvalue equation (\ref{eq:eig_vector}) with eigenvalue $\lambda^2$.  This vector generates a symmetric traceless tensor of the form
\begin{equation}
	H_{\mu\nu} = \nabla_{\{\mu} \xi_{\nu\}} \equiv \nabla_\mu \xi_\nu + \nabla_\nu \xi_\mu - g_{\mu\nu} \nabla^\rho \xi_\rho~.
\label{eq:pure_diff}
\end{equation}
It is then straightforward to apply the Laplacian to this expression and show that $H_{\mu\nu}$ satisfies its own eigenvalue equation \eqref{eq:eig_tensor} with the same value of $\lambda^2$. This proves the claim in one direction, i.e. that the spectrum of a normalizable vector field maps onto the spectrum of a symmetric traceless tensor.

To prove the opposite direction and explain the exceptions, we first need to ask whether all symmetric tensors can be presented as diffeomorphisms.  Let's assume the contrary, that there is a normalizable mode $\mathcal{H}_{\mu\nu}$ that cannot be written as in (\ref{eq:pure_diff}) for any normalizable vector $\xi_\mu$.  Then its contraction with the expression on the right-hand side of (\ref{eq:pure_diff}) must vanish for any $\xi_\mu$.  We can then integrate this contraction over the entire geometry. Because the vector and tensor fields must both be normalizable the integral only vanishes if $(\nabla^\mu \mathcal{H}_{\mu\nu})\xi^\nu = 0$ for all vectors $\xi^\nu$.  This is only possible if the tensor is transverse $\nabla^\mu \mathcal{H}_{\mu\nu} = 0$.

The upshot of this discussion is that the eigenvalue equation for a tensor field on AdS$_2$ (\ref{eq:eig_tensor}) can be solved exactly by a Hodge decomposition of the form
\begin{equation}
	H_{\mu\nu} = \nabla_{\{\mu} \xi_{\nu\}} + \mathcal{H}_{\mu\nu}~,
\end{equation}
where $\xi_\mu$ satisfies the vector eigenvalue equation (\ref{eq:eig_vector}) with the same value of $\lambda^2$ and $\mathcal{H}_{\mu\nu}$ is transverse.  This transversality condition places two conditions on the component fields of $\mathcal{H}_{\mu\nu}$ and is quite restrictive.  In holomorphic coordinates, these conditions take the form $\nabla_{\bz} \mathcal{H}_{zz} = \partial_{\bz} \mathcal{H}_{zz} = 0$, plus the holomorphic conjugate.  The solutions to this are
\begin{equation}
	\mathcal{H}_{zz}^{(n)} = \ell^2\sqrt{\frac{n(n^2-1)}{2\pi}} z^{n-2}~,
\label{eq:tensor_basis}
\end{equation}
where $n$ is an integer that takes the values $2,3,\ldots$.  These field configurations are \emph{quadratic holomorphic differentials}.  Importantly, they are normalizable modes, with the overall normalization above chosen such that $ \langle \mathcal{H}^{(m)} | \mathcal{H}^{(n)} \rangle = \ell^2 \delta_{m,n}$.  Additionally, it is straightforward to show that the quadratic holomorphic differentials satisfy the eigenvalue equation
\begin{equation}
	\Delta^{(2)} \mathcal{H}_{\mu\nu}^{(n)} = -\frac{2}{\ell^2} \mathcal{H}^{(n)}_{\mu\nu}~.	
\end{equation}
In other words, these modes are eigenfunctions of the tensor kinetic operator with eigenvalues $\lambda^2= - \frac{2}{\ell^2}$.

Physically, the quadratic holomorphic differentials are manifestations of the symmetries of AdS$_2$.  There are an infinite number of conformal Killing vectors on AdS$_2$, as discussed in subsection~\ref{sec:jt:euclid}, enumerated by $n \in \mathbb{Z}$.  These vector fields satisfy the Euclidean eigenvalue equation (\ref{eq:eig_vector}) with eigenvalue $\lambda^2 = - \frac{2}{\ell^2}$.  They are not normalizable, though, and so they are not physical vector modes.  However, when we dualize these conformal Killing vectors to tensor fields, some of the corresponding tensor modes are normalizable.  These normalizable tensor modes are precisely the quadratic holomorphic differentials (\ref{eq:tensor_basis}).  The conformal Killing vectors on AdS$_2$ therefore generate physical tensor modes that must be summed over when we quantize a gravitational theory living on AdS$_2$.

In summary, we have established that any vector field can be dualized to two scalars plus a harmonic vector field.  Similarly, a symmetric traceless tensor field can be dualized to a vector plus a transverse tensor, with the latter corresponding to the quadratic holomorphic differentials. Therefore, the tensor can be further dualized to two scalars, a harmonic vector, and the quadratic holomorphic differentials.  All the scalar fields will have a {\it continuous} spectrum, while the harmonic vectors and transverse tensors have a \emph{discrete} spectrum.  These discrete modes decouple from the continuous modes, but they must be included in the full one-loop partition function.

\subsection{\texorpdfstring{AdS$_2$}{AdS2} as a Worldsheet}
\label{sec:quant:worldsheet}

The arguments and results presented in the preceding subsection are similar to standard ones in critical bosonic string theory, but they are not identical. For example, in our spacetime context we analyze a symmetric tensor $H_{\mu\nu}$ that we take to be traceless, but there is no underlying Weyl symmetry that forces it to be traceless. It is instructive to compare the two situations in the formalism that is familiar from bosonic string theory.

In the textbook version of gauge-fixed worldsheet string theory we must pay special attention to {\it residual} diffeomorphisms that 
can be exploited to fix some of the vertex operators. Their number is counted by the conformal Killing vectors which in turn are the normalizable 
zero modes of the operator $P_1$ defined by
\begin{equation}
	(P_1 \xi)_{\mu\nu} = \nabla_\mu \xi_\nu + \nabla_\nu \xi_\mu - g_{\mu\nu}\nabla^\rho \xi_\rho~.
\end{equation}
On the other hand, gauge-fixing of worldsheet diffeomorphisms also leaves unfixed {\it moduli} that must be integrated over explicitly. 
Their number is counted by the quadratic holomorphic differentials which in turn are the normalizable zero modes of the adjoint operator $P_1^T$.  
An important constraint on these numbers is the Riemann-Roch theorem that yields their difference
\begin{equation}
	\text{Ker}\,P_1 - \text{Ker}\,P_1^T = 3 \chi~,
\end{equation}
where the Euler characteristic $\chi$ is given by the Gauss-Bonnet theorem
\begin{equation}
	\chi = \frac{1}{4\pi}\left( \int d^2x\,\sqrt{g}\,R + 2 \int dt\,\sqrt{\gamma}\,K\right)~.
\label{eq:gaussbonnet}
\end{equation}

The sphere $S^2$ has $\chi = 2$ and satisfies the Riemann-Roch theorem with $\text{Ker}\,P_1 = 6$ real conformal Killing vectors and $\text{Ker}\,P_1^T = 0$ quadratic holomorphic differentials.  The disk $D^2$ has $\chi = 1$ and we can think of it as the sphere $S^2$ with holomorphic and anti-holomorphic coordinates identified such that the Riemann-Roch theorem holds with $\text{Ker}\,P_1 = 3$ real conformal Killing vectors and $\text{Ker}\,P_1^T = 0$.

The AdS$_2$ geometry is the hyperbolic plane $H^2$ and is related to the disk $D^2$ via a Weyl transformation that diverges on the boundary of the disk.  In holographic applications we usually interpret the AdS$_2$ geometry as the limit of regularized AdS$_2$, where the regulating boundary is removed by a cut-off surface which is subsequently taken towards the asymptotic boundary.  The Euler characteristic remains $\chi = 1$ for each regularized disk and so $\chi = 1$ holds also for the AdS$_2$ limit.

This result is realized in the Gauss-Bonnet theorem (\ref{eq:gaussbonnet}) as follows.  The curvature is constant, $R = -\frac{2}{\ell^2}$, and gets multiplied by the overall divergent volume.  However, the boundary term cancels the divergent bulk volume while also adding a finite, negative term to the on-shell action.  This is precisely the well-known mechanism that renormalizes the AdS$_2$ volume to $-2\pi \ell^2$. The Euler characteristic  $\chi = 1$ of $H^2$ follows from multiplication of the curvature $R = -\frac{2}{\ell^2}$ by this renormalized, negative volume.

Importantly, though, while AdS$_2$ has the same genus as a disk, they realize the Riemann-Roch theorem differently. As stressed in subsection~\ref{sec:jt:euclid}, there are no \emph{normalizable} conformal Killing vectors in AdS$_2$.  But, as we established in subsection~\ref{sec:quant:dual}, there is an infinite tower of normalizable quadratic holomorphic differentials.  At any point in AdS$_2$, we can sum over this tower to find a constant local density of modes:  
\begin{equation}\begin{aligned}
	\sum_{n=2}^\infty \left( |\mathcal{H}_{zz}^{(n)}|^2 + \text{h.c.} \right) &= \sum_{n=2}^\infty \frac{n(n^2-1)}{4\pi \ell^2} |z|^{2(n-2)} \left(1- |z|^2\right)^4 = \frac{3}{2\pi \ell^2}~.
\end{aligned}\end{equation}
The regularized AdS$_2$ volume is $-2\pi \ell^2$, and so (after regularization) the number of quadratic holomorphic differentials comes out to be
\begin{equation}
	\text{Ker}\,P_1^T = -2\pi\ell^2 \times \frac{3}{2\pi \ell^2} = -3~.
\end{equation}
We therefore find that the Riemann-Roch theorem is satisfied in AdS$_2$ spacetimes with $\text{Ker}\,P_1 = 0$ real conformal Killing vectors and $\text{Ker}\,P_1^T = -3$ quadratic holomorphic differentials.

The fundamental difference between these two perspectives comes from considerations of Weyl symmetry.  In worldsheet string theory the Weyl symmetry is exact and must be gauge-fixed to eliminate redundancies.  Gravitational theories on nearly-AdS$_2$ spacetimes, on the other hand, have a classical Weyl symmetry that is broken by the vacuum.  The Goldstone bosons reflecting this symmetry breaking are the quadratic holomorphic differentials which are therefore physical degrees of freedom.

\subsection{Free Field Partition Function}

In preparation for computing the partition function of the JT model in the next section, we now address the closely-related problem where each field in the model is 
replaced by its minimally-coupled analogue and propagates on a fixed AdS$_2$ background.  

The graviton can be decomposed into its scalar trace plus a symmetric traceless tensor, while the dilaton is simply a real scalar field. Furthermore, diffeomorphism-invariance of the theory acts as a gauge symmetry on these fields that requires the addition of two real vector Fadeev-Popov ghosts\footnote{We discuss the gauge-fixing procedure in subsection \ref{sec:quadact}. Two vector ghosts are needed in harmonic gauge.}, in order to pick out a particular gauge orbit.
We thus consider a symmetric traceless tensor, two scalars, and two vector ghosts. In this subsection we further endow all fields with common mass $m$. The one-loop partition function will then be given by
\begin{equation}
	Z = \sqrt{\frac{\text{det}\,(\Delta^{(1)} + m^2 )^2}{\text{det}\,(\Delta^{(0)} + m^2)^2\,\,\text{det}\,(\Delta^{(2)} + m^2)}}~,
\label{eq:part_full}
\end{equation}
where $\Delta^{(0)}$, $\Delta^{(1)}$, and $\Delta^{(2)}$ are the kinetic operators for massless scalars, vectors, and tensors that 
were introduced in subsections \ref{sec:quant:clas} and \ref{sec:quant:dual}. The ghost fields have anti-commuting statistics, and so their contribution to the partition function (\ref{eq:part_full}) is in the numerator rather than in the denominator.

Each of these functional determinants can be evaluated explicitly on their own.  For example, 
the functional determinant for a single scalar kinetic operator will involve explicitly summing over all eigenvalues of the continuous modes $u_{p m}$:
\begin{equation}
	\log Z = -\frac{1}{2}\log \det\left(\Delta^{(0)} + m^2\right) = -\frac{1}{2}\int_{-\infty}^{+\infty} dp\, \mu(p) \left(p^2 + \frac{1}{4}+ m^2\ell^2\right)~,
\end{equation}
with the spectral density $\mu(p)$ given in \eqref{eqn:Planch}.
As is standard in one-loop calculations in quantum field theory, this expression is formally divergent and requires careful regularization.  However, in our current setup, we do not actually need to evaluate this quantity.  As far as the continuous sector is concerned, each of the real vector ghosts can be dualized to two scalar fields, while the symmetric traceless tensor can be dualized to a vector (which is then further dualized to two scalars).  Therefore the net spectrum consists of four scalar fields and four ghost scalars, all with exactly the same spectrum of eigenvalues.  We therefore have a precise cancellation between the physical and ghost modes in the continuous sector, leaving us with
\begin{equation}
	Z_\text{cont} = 1~.
\end{equation}

However, as emphasized in subsection~\ref{sec:quant:dual}, it is not possible to dualize vectors and tensors entirely to scalars; we must also carefully consider the discrete mode sector.  
We first consider the harmonic vectors given in (\ref{eq:harmonic_basis}) as $\mathcal{A}_z^{(n)}$ with $n = 1 , 2,\ldots$ (as well as their Hermitian conjugates).
They appear in the Hodge decompositions of the two vector ghosts and also in the symmetric traceless tensor, effectively yielding a single tower of discrete modes from one real ghost. The kinetic operator $\Delta^{(1)}$ vanishes on the harmonic vectors so the  operators  $\Delta^{(1)}+m^2$ in the determinant  \eqref{eq:part_full} have eigenvalue $m^2$.  The contribution to the partition function from harmonic vectors is therefore given by
\begin{equation}
	Z_\text{harmonic} = \prod_{n=1}^\infty \left(m^2 \ell^2\right)~.
\end{equation}
We also need to consider the quadratic holomorphic differentials given in (\ref{eq:tensor_basis}) as $\mathcal{H}^{(n)}_{zz}$ with $n = 2,3,\ldots$ (as well as their Hermitian conjugates). They appear in the decomposition of the symmetric traceless tensor and have eigenvalue $m^2\ell^2 - 2$. Their contribution to the partition function is
\begin{equation}
	Z_\text{QHD} = \prod_{n=2}^\infty \left(m^2 \ell^2 - 2\right)~.
\end{equation}

There is no mixing between the continuous modes, the discrete vector modes, and the discrete QHDs in the path integral. Therefore, the full one-loop partition function
\eqref{eq:part_full} becomes
\begin{equation}\begin{aligned}
	\log Z &= \log Z_\text{cont} + \log Z_\text{harmonic} + \log Z_\text{QHD} \\
	&= \sum_{n=1}^\infty \log m^2 \ell^2 - \sum_{n=2}^\infty \log \left(m^2 \ell^2 - 2 \right)~.
\end{aligned}\end{equation}
Of course, this simpler expression is still obviously UV-divergent.  There are many ways to regularize it, such as Pauli-Villars regularization or heat kernel regularization. The simplest way to do so in this free-field example is through zeta-function regularization which easily gives 
\begin{equation}\begin{aligned}
	\log Z &= \zeta(0)\log\,m^2 \ell^2 - (\zeta(0) - 1)\log\left(m^2 \ell^2 - 2\right) \\
	&= -\frac{1}{2}\log\,m^2 \ell^2 + \frac{3}{2}\log\left(m^2 \ell^2 - 2\right)~.
\label{eq:znont}
\end{aligned}\end{equation}
The removal of the UV-divergence leaves ambiguous the finite coefficients of power-law corrections, but \eqref{eq:znont} comprises all terms in $Z$ that depend logarithmically on $m$. 

The result \eqref{eq:znont} is finite only when $m^2 > \frac{2}{\ell^2}$. In the next section we will see that the JT model is similar to the free model where all fields have an on-shell mass of $m^2 = \frac{2}{\ell^2}$ due to their coupling to the cosmological constant. Then the first term in \eqref{eq:znont} is a finite constant while the second is logarithmically divergent. After also taking the coupling to the background dilaton properly into account, the latter will essentially become the logarithm of the small dilaton slope. 

\section{Quantum Corrections to the Jackiw-Teitelboim Model}
\label{sec:oneloop}

In this section we compute the logarithmic corrections to the one-loop partition function in  the Jackiw-Teitelboim Model. We first determine the gauged-fixed form of the action for quadratic fluctuations around the black hole background and then compute the corresponding functional determinant perturbatively away from the free theory discussed in the previous section. We exhibit several distinct cancellations before concluding that the quantum corrections are dominated by quadratic holomorphic differentials (QHDs) perturbed by the dilaton. 

\subsection{Effective Field Theory Expectations}
\label{sec:EFTE}
The free field modes considered in section~\ref{sec:quant} diagonalize their canonical Laplacians with the requisite spin. 
The true gauged-fixed quadratic action that we determine in this section will be different in detail so the free modes will not be eigenfunctions
but they will nonetheless form a complete basis for all quantum fluctuations.  

The non-minimal couplings will present several challenges but ultimately we will find that the exact cancellations in the free field partition function 
will carry over to the complete Jackiw-Teitelboim model, albeit approximately, due to the mild conformal symmetry breaking.  These approximate cancellations will be enough to show that the logarithmic terms in the partition function are entirely due to the quadratic holomorphic differentials.

\begin{figure}[ht]\centering
	\includegraphics[scale = 0.8]{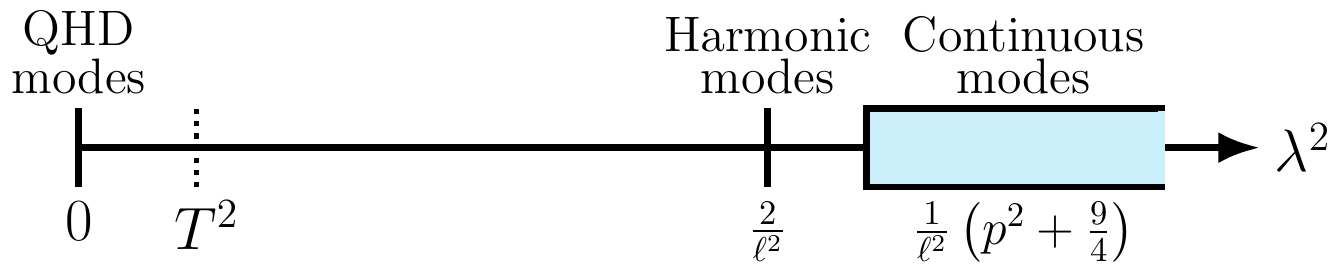}
	\caption{The spectrum of various types of Euclidean AdS$_2$ field modes with an on-shell mass of $m^2 = \frac{2}{\ell^2}$.  The quadratic holomorphic differentials are zero modes in the extremal limit, so they dominate finite-temperature effects.}
	\label{fig:scales2}
\end{figure}

Before getting into technical details, it is worth outlining why this result is expected from effective field theory,  as depicted in figure~\ref{fig:scales2}. Intuitively, we will interpret all fields in the JT model as scalars with on-shell mass $m^2 = \frac{2}{\ell^2}$, which arises from their coupling to the background cosmological constant. 
In AdS$_2$ their continuous off-shell (Euclidean) spectrum is strictly larger with $\lambda^2\geq \frac{9}{4\ell^2}$. In this terminology zero-modes (a.k.a. harmonic modes) have $\lambda^2=\frac{2}{\ell^2}$ and the ``true" zero-modes with $\lambda^2=0$ are the QHDs. 

Thermal effects due to the black hole background modify all these values but,
in the near-extremal regime we consider, the energy scale $\sim \frac{1}{\ell^2}$ is parametrically larger than the scale set by the black hole temperature, and so the finite-temperature effects will generally be subleading.  However, the partition function of the QHDs will be dominated by the thermal effects. We interpret them 
as near-zero modes that are lifted above zero by the conformal symmetry breaking and give correspondingly large contributions to the Euclidean partition function of the theory.

\subsection{One-Loop Quadratic Action}
\label{sec:quadact}

We expand the dilaton and the metric to quadratic order around their classical background values via the variations
\begin{equation}
	\delta g_{\mu\nu} = h_{\mu\nu}~, \quad \delta \Phi = \phi~.
\end{equation}
It is useful to decompose the metric fluctuations into its trace and its traceless components, given respectively by
\begin{equation}
	h = h\ind{_\mu^\mu}~, \quad \bar{h}_{\mu\nu} = h_{\mu\nu} - \frac{1}{2}g_{\mu\nu} h~.
\end{equation}
Using the background equations of motion, the second variation of the action (\ref{eq:sjt}) can then be put in the form
\begin{equation}\begin{aligned}
	\delta^2 I_\text{JT} &= -\frac{1}{16\pi G_2}\int d^2x\,\sqrt{g}\,\bigg{[} \frac{1}{2}\bar{h}_{\mu\nu}\left(\square+\frac{2}{\ell^2}\right) \bar{h}^{\mu\nu} + (\nabla^\mu \bar{h}_{\mu\nu})(\nabla_\rho \bar{h}^{\rho\nu})  \\
	&\quad + 2 \phi \nabla^\mu \nabla^\nu \bar{h}_{\mu\nu}- \phi \left(\square-\frac{2}{\ell^2}\right) h + (\nabla^\mu \Phi)\left(2 \bar{h}_{\mu\nu}\nabla_\rho \bar{h}^{\rho\nu} + h \nabla^\nu \bar{h}_{\mu\nu}\right) \\
	&\quad + (\Phi - 1)\left(\frac{1}{2}\bar{h}_{\mu\nu}\left(\square+\frac{1}{\ell^2}\right) \bar{h}^{\mu\nu} + (\nabla^\mu \bar{h}_{\mu\nu})(\nabla_\rho \bar{h}^{\rho\nu}) + \frac{h^2}{2\ell^2}\right) \bigg{]}~.
\label{eq:d2s}
\end{aligned}\end{equation}
We chose Dirichlet boundary conditions for the field variations that force $h_{\mu\nu}$ and $\phi$ to vanish on the boundary, and so boundary terms in the action \eqref{eq:d2s} are absent. 

We fix the gauge of the metric fluctuations to the harmonic gauge
\begin{equation}
	G_\mu \equiv \nabla^\nu \bar{h}_{\mu\nu}=0~,
	\label{eqn:harmonic}
\end{equation}
by adding the gauge-fixing action
\begin{equation}
	I_\text{g.f.} = \frac{1}{16\pi G_2}\int d^2x\,\sqrt{g}\, \xi^{-1}\,G_\mu G^\mu~, 
\label{eq:sgf}
\end{equation}
where $\xi$ is an arbitrary numerical gauge parameter.  Under an infinitesimal diffeomorphism $x^\mu \to x^\mu + \xi^\mu$, the field fluctuations transform as $\delta h_{\mu\nu} = 2 \nabla_{(\mu}\xi_{\nu)}$. The corresponding functional derivative of the gauge-fixing function $G_\mu$ is then given by
\begin{equation}
	\frac{\delta G_{\mu}}{\delta \xi^\nu} = g_{\mu\nu}\left(\square -\frac{1}{\ell^2}\right)~.
\end{equation}
The functional determinant of this variation appears in the path-integral when using the Faddeev-Popov procedure to impose the harmonic gauge 
condition (\ref{eqn:harmonic}) at the quantum level. We include it by introducing anti-commuting vector ghosts $b_\mu,c_\mu$ with action
\begin{equation}
	I_\text{ghost} = -\frac{1}{16\pi G_2}\int d^2x\,\sqrt{g}\, b_\mu\left(\square -\frac{1}{\ell^2} \right) c^\mu~.
\label{eq:sgh}
\end{equation}
That the ghosts are vectors with a second order action is due to the harmonic gauge condition and follows algorithmically from the steps above. Moreover, we devised the particular gauge function so that these two vector ghosts are free. Other choices could introduce couplings between the ghosts and the background dilaton which would be technically more complicated without changing any physical observables.

We now add the gauge-fixing term (\ref{eq:sgf}) and the ghost action (\ref{eq:sgh}) to the quadratic variation (\ref{eq:d2s}) of the original JT action and find the total one-loop quadratic action
\begin{equation}\begin{aligned}
	I &= -\frac{1}{16\pi G_2}\int d^2x\,\sqrt{g}\,\bigg{[} \frac{1}{2}\bar{h}_{\mu\nu}\left(\square+\frac{2}{\ell^2}\right) \bar{h}^{\mu\nu} + \frac{\xi - 1}{\xi}(\nabla^\mu \bar{h}_{\mu\nu})(\nabla_\rho \bar{h}^{\rho\nu}) \\
	&\quad + 2 \phi \nabla^\mu \nabla^\nu \bar{h}_{\mu\nu} - \phi \left(\square-\frac{2}{\ell^2}\right) h + (\nabla^\mu \Phi)\left(2 \bar{h}_{\mu\nu}\nabla_\rho \bar{h}^{\rho\nu} + h \nabla^\nu \bar{h}_{\mu\nu}\right) \\
	&\quad + (\Phi - 1)\left(\frac{1}{2}\bar{h}_{\mu\nu}\left(\square+\frac{1}{\ell^2}\right) \bar{h}^{\mu\nu} + (\nabla^\mu \bar{h}_{\mu\nu})(\nabla_\rho \bar{h}^{\rho\nu}) + \frac{h^2}{2\ell^2}\right) \\
	&\quad + b_\mu \left(\square - \frac{1}{\ell^2}\right)c^\mu \bigg{]}~.
\label{eq:squad}
\end{aligned}\end{equation}
The corresponding eigenvalue equations for the field fluctuations are
\begin{equation}\begin{aligned}
	\left(-\square - \frac{2}{\ell^2} - \lambda^2\right) \bar{h}_{\mu\nu} &= \frac{2(1-\xi)}{\xi} \nabla_{\{\mu} \nabla^\rho \bar{h}_{\nu\}\rho} +2\nabla_{\{ \mu} \nabla_{\nu \}} \phi \\
	&\quad + (\Phi-1)\left( \square \bar{h}_{\mu\nu} - 2\nabla_{\{\mu} \nabla^\rho \bar{h}_{\nu\}\rho} \right) \\
	&\quad + (\nabla^\rho \Phi)\left(\nabla_\rho \bar{h}_{\mu\nu} - 2\nabla_{\{\mu}\bar{h}_{\nu\}\rho} - g_{\rho\{\mu} \nabla_{\nu\}} h \right)~, \\
	\left(-\square + \frac{2}{\ell^2} - \lambda^2\right) h &= -2 \nabla^\mu \nabla^\nu \bar{h}_{\mu\nu}~,\\
	\left(-\square + \frac{2}{\ell^2} - \lambda^2\right) \phi &= -(\Phi-1)\frac{h}{\ell^2} - (\nabla^\mu \Phi) \nabla^\nu \bar{h}_{\mu\nu}~, \\
	\left(-\square + \frac{1}{\ell^2} - \lambda^2 \right) c_\mu &= \left(-\square + \frac{1}{\ell^2} - \lambda^2\right) b_\mu = 0~,
\label{eq:quad_eom}
\end{aligned}\end{equation}
where we have defined $\mathcal{O}_{\{\mu\nu\}} \equiv \frac{1}{2}\left(\mathcal{O}_{\mu\nu} + \mathcal{O}_{\nu\mu} - g_{\mu\nu} g^{\rho\sigma}\mathcal{O}_{\rho\sigma}\right)$ as the symmetrized, traceless part of any tensor $\mathcal{O}_{\mu\nu}$.  

The kinetic terms on the left-hand side of the eigenvalue equations correspond for all fields to an on-shell mass $m = \frac{2}{\ell^2}$ (where by ``mass'' we refer to the scalar mass after appropriate dualizations), equivalent to conformal dimension $h = 2$. Therefore these kinetic terms by themselves define the free model we analyzed in section~\ref{sec:quant}, with this specific value of the mass. 

The right-hand sides of the eigenvalue equations simplify considerably in the extremal limit where the background dilaton has no slope and we can simply set $\Phi = 1$. Even in this limit, though, they represent non-trivial kinetic terms beyond those of the free model in section~\ref{sec:quant}. Moreover, once the dilaton slope gets turned on, the problem becomes much harder because the coupling to the background dilaton is fairly complicated. In the following subsections we compute the one-loop action by first expanding on the free field basis and then computing perturbatively in powers of $\sqrt{\mu a}$, the small parameter that controls the dilaton profile.

\subsection{Continuous Modes}
\label{sec:oneloop:cont}

In this subsection we address the  contributions from continuous modes to the partition function. They come from dualizing all fields to scalars, and then expanding these scalars on the complete basis of functions $\{ u_{p m}\}$ that we introduced in subsection \ref{sec:quant:clas}.  Recall that these functions are defined for all radial momenta $p \in \mathbb{R}$ and all azimuthal quantum numbers $m \in \mathbb{Z}$, and that they are normalized such that
\begin{equation}
	\langle u_{p m} | u_{p' m'}\rangle = \int d^2x\, \sqrt{g}\, u_{p' m'} u_{p m} = \ell^2 \delta(p - p') \delta_{m,m'}~.
\end{equation}
When we dualize a vector to scalars, we get two distinct sets of continuous modes given by
\begin{equation}
	(v^{\parallel}_{p m})_{\mu} = \frac{\ell}{\sqrt{p^2 + \frac{1}{4}}}\nabla_\mu u_{p m}~, \quad (v^{\perp}_{p m})_\mu = \frac{\ell}{\sqrt{p^2 + \frac{1}{4}}}\epsilon_{\mu\nu}\nabla^\nu u_{p m}~.
\label{eq:vcont}
\end{equation}
Similarly, when we dualize the traceless symmetric tensor to scalars, we get two more distinct sets of continuous modes:
\begin{equation}\begin{aligned}
	(w^{\parallel}_{p m})_{\mu\nu} &= \frac{\ell^2}{\sqrt{2\left(p^2 + \frac{1}{4}\right) \left(p^2 + \frac{9}{4}\right)}} \nabla_{\{\mu} \nabla_{\nu\}} u_{p m}~, \\
	(w^{\perp}_{p m})_{\mu\nu} &= \frac{\ell^2}{\sqrt{2\left(p^2 + \frac{1}{4}\right) \left(p^2 + \frac{9}{4}\right)}} \nabla_{\{\mu} \epsilon_{\nu\}\rho}\nabla^\rho u_{p m}~.
\label{eq:tcont}
\end{aligned}\end{equation}
The prefactors in (\ref{eq:vcont}) and (\ref{eq:tcont}) for the vector and tensor modes are chosen such that they satisfy the normalization conditions
\begin{equation}\begin{aligned}
	&\langle v_{p' m'}^{\parallel} | v_{p  m}^{\parallel}\rangle = \langle v_{p' m'}^{\perp} | v_{p m}^{\perp}\rangle = \ell^2 \delta(p - p') \delta_{m, m'}~, \quad &\langle v_{p' m'}^{\parallel} | v_{p m}^{\perp}\rangle = 0~,  \\
	&\langle w_{p' m'}^{\parallel} | w_{p m}^{\parallel}\rangle = \langle w_{p' m'}^{\perp} | w_{p m}^{\perp}\rangle = \ell^2\delta(p - p')\delta_{m,m'}~, \quad &\langle w_{p' m'}^{\parallel} | w_{p  m}^{\perp}\rangle = 0~.
\end{aligned}\end{equation}

We expand all fluctuating fields in our quadratic action on these bases: 
\begin{equation}\begin{aligned}
	&\bar{h}_{\mu\nu} = \sum_{m}\int dp\, (c_{1p m} w^{\parallel}_{p m} + c_{2p m} w^{\perp}_{p m})_{\mu\nu}~, \\
	&h = \sum_{m}\int dp\, c_{3p m} u_{p m}~, \quad \phi = \sum_{m}\int dp\, c_{4p m} u_{p m}~, \\
	&b_\mu = \sum_{m}\int dp\, (c_{5p m} v^{\parallel}_{p m} + c_{6p m} v^{\perp}_{p m})_{\mu}~, \\
	&c_\mu = \sum_{m}\int dp\, (c_{7p m} v^{\parallel}_{p m} + c_{8p m} v^{\perp}_{p m})_{\mu}~,
\label{eq:cont_exp}
\end{aligned}\end{equation}
with arbitrary constants $c_{ip m}$ parametrizing the configuration space of all fields. 

We now need to evaluate the (Euclideanized) quadratic action (\ref{eq:squad}) over these expansions of our fields in order to compute the continuous mode contribution to the one-loop partition function\footnote{When we Euclideanize the one-loop quadratic action, we also need to Wick-rotate the scalar metric fluctuation $h \to i h$ such that its kinetic term becomes positive-definite.  This is the standard resolution to the well-known conformal factor problem in Euclidean quantum gravity~\cite{Gibbons:1978ac,Schleich:1987fm,Mazur:1989by}. This procedure is the origin of explicit factors of $i$ in the matrices.}.  The physical fields decouple from the ghost fields so we can decompose the action as
\begin{equation}
	I = I_b + I_\text{ghost}~,
\end{equation}
where $I_b$ is the action for the physical bosonic fields and $I_\text{ghost}$ the action for the ghost fields. Upon expansion on our basis modes these contributions become
\begin{equation}\begin{aligned}
	I_b &= \sum_{m,m'} \int dp\,dp'\,
	\begin{pmatrix}
		c_{1 p m} & \ldots & c_{4p m}	
	\end{pmatrix}
	M_{p m, p' m'}
	\begin{pmatrix}
		c_{1 p' m'} \\
		\vdots \\
		c_{4 p' m'}
	\end{pmatrix}~, \\
	I_\text{ghost} &= \sum_{m,m'} \int dp\,dp'\,
	\begin{pmatrix}
		c_{5 p m} & \ldots & c_{8 p m}	
	\end{pmatrix}
	N_{p m, p' m'}
	\begin{pmatrix}
		c_{5 p' m'} \\
		\vdots \\
		c_{8 p' m'}
	\end{pmatrix}~,
\end{aligned}\end{equation}
where $M$, $N$ are $4 \times 4$ matrices with indices ranging over all quantum numbers.  We further decompose 
the matrix $M$ as
\begin{equation}
	M_{pm,p'm'} = M_{pm,p'm'}^{(0)} + M_{pm,p'm'}^{(1)}~,
\end{equation}
where $M^{(0)}$ is the zero-temperature piece that comes from setting the dilaton to the constant $\Phi = 1$, while $M^{(1)}$ is the additional finite-temperature contribution that depends on the full dilaton profile.  Explicit computation of these matrices gives
\begin{equation}
	M^{(0)}_{p m,p'm'} = \frac{p^2 + \frac{9}{4}}{32\pi G_2} \delta(p-p')\delta_{m,m'}
	\begin{pmatrix} 
		\xi^{-1} & 0 & 0 & - \sqrt{\frac{2 p^2 + \frac{1}{2}}{p^2 + \frac{9}{4}}} \\
		0 & \xi^{-1} & 0 & 0 \\
		0 & 0 & 0 & -i \\
		- \sqrt{\frac{2 p^2 + \frac{1}{2}}{p^2 + \frac{9}{4}}} & 0 & -i & 0
	\end{pmatrix}~,
\label{eq:m0}
\end{equation}
\begin{equation}
	M^{(1)}_{p m, p' m'} = \frac{1}{32\pi G_2}\begin{pmatrix}
	0 & 0 & i \sqrt{\frac{p^2 + \frac{9}{4}}{p^2 + \frac{1}{4}}} \mathcal{R}_{p' m', p m} & 0 \\
	0 & 0 & i \sqrt{\frac{p^2 + \frac{9}{4}}{p^2 + \frac{1}{4}}} \mathcal{L}_{p' m', p m} & 0 \\
	i \sqrt{\frac{{p'}^2 + \frac{9}{4}}{{p'}^2 + \frac{1}{4}}}\mathcal{R}_{p m, p' m'} & i \sqrt{\frac{{p'}^2 + \frac{9}{4}}{{p'}^2 + \frac{1}{4}}}\mathcal{L}_{p m, p' m'} & -\mathcal{R}_{(p m, p' m')} & 0 \\
	0 & 0 & 0 & 0
	\end{pmatrix}~,
\label{eq:m1}
\end{equation}
where the functions $\mathcal{L}$ and $\mathcal{R}$ are defined as the integrals over the dilaton profile and the scalar wavefunctions:
\begin{equation}\begin{aligned}
	\mathcal{L}_{p m,p' m'} &\equiv \int d^2x\,\sqrt{g}\,\epsilon^{\mu\nu}\left(\nabla_\nu \Phi\right) u_{pm} \nabla_\mu u_{p'm'}~, \\
	\mathcal{R}_{p m,p' m'} &\equiv \int d^2x\,\sqrt{g}\,\left(\nabla^\mu \Phi\right) u_{pm} \nabla_\mu u_{p'm'}~.
\label{eq:mlr}
\end{aligned}\end{equation}
The matrix $M^{(0)}$ is relatively simple because it does not mix different values of $p$ nor $m$. The only mild complication is the off-diagonal terms due to 
the kinetic term $\phi \nabla^\mu \nabla^\nu \bar{h}_{\mu\nu}$ that appears in the action for quadratic fluctuations in the Jackiw-Teitelboim model but not in our free benchmark model. 

On the other hand, the matrix $M^{(1)}$ is complicated because it mixes different values of $p$ and $m$ and depends on the dilaton profile.  Crucially, though, most of its entries are zero.  This is not obvious; it is the result of an intricate cancellation among the various terms in the action (\ref{eq:squad}) when we evaluate its matrix elements
on the basis of continuous field modes.

The final matrix is the ghost matrix $N$. It is nearly trivial, because we intentionally chose a gauge-fixing condition that simply gave two free vector ghosts.  Explicitly, $N$ is given by
\begin{equation}
	N_{p m,p'm'} = \frac{p^2 + \frac{9}{4}}{32\pi G_2} \delta(p-p')\delta_{m,m'}
	\begin{pmatrix} 
		0 & -1 & 0 & 0 \\
		-1 & 0 & 0 & 0 \\
		0 & 0 & 0 & -1 \\
		0 & 0 & -1 & 0
	\end{pmatrix}~.
\end{equation}
As expected, this matrix does not mix different values of $(p,m)$ because the ghost fields are minimally-coupled to the background metric and do not interact with the background dilaton.

The path integral in the continuous mode sector at this point has reduced to an ordinary Gaussian integral over the coefficients $c_{i p m}$. We find
\begin{equation}
	Z_\text{cont} = Z_b Z_\text{ghost}~,
\end{equation}
where 
\begin{equation}
	Z_{b} = \frac{\prod_{p,m}\pi^2}{\sqrt{\det M}}~, \quad Z_\text{ghost}^{-1} =\frac{\prod_{p,m}\pi^2}{\sqrt{\text{det}\,N}}~.
\end{equation}
Each determinant is over the quantum numbers $(p,m)$ as well as the $4 \times 4$ matrices themselves. The ghosts contribute as an inverse due to their Fermi statistics. 

The requisite functional determinants are difficult to compute in general but they are simple in the extremal limit $\sqrt{\mu a} \to 0$. 
Accordingly, we temporarily ignore the finite-temperature piece $M^{(1)}$ and then find the partition functions
\begin{equation}\begin{aligned}
	Z_b(T = 0) &= \prod_{p ,m}\left(\frac{32 \pi^2 G_2}{p^2 + \frac{9}{4}}\right)^2\xi~, \\
	Z_\text{ghost}^{-1}(T = 0) &= \prod_{p,m}\left(\frac{32 \pi^2 G_2}{p^2 + \frac{9}{4}}\right)^2~.
\label{eq:zcont_phys,eq:zcont_ghost}
\end{aligned}\end{equation}
Despite the couplings between the metric and dilaton fluctuations in the one-loop quadratic action, encoded in the real off-diagonal terms of $M^{(0)}$ displayed in \eqref{eq:m0}, the results \eqref{eq:zcont_phys,eq:zcont_ghost} are precisely what we would obtain for entirely free fields.  Moreover, all physical results are independent of the gauge-fixing parameter
so we can take  $\xi = 1$ without loss of generality. With this choice the physical and ghost contributions to the continuous sector partition function $Z_\text{cont}$ in the extremal limit are manifestly inverses of one another and therefore cancel.  We are left with 
\begin{equation}
	Z_\text{cont}(T = 0) = 1~,
\end{equation} 
with no regularization required.  This is consistent with prior results~\cite{Grumiller:2015vaa} finding that, for a constant dilaton, there are no perturbative quantum corrections to the classical partition function.  

We now need to address the finite-temperature case. In the low-temperature regime we can make progress in perturbation theory, expanding the determinant of $M$ as:
\begin{equation}
	\det M = \det M^{(0)}\left(1 + \text{tr}\,M^{(1)} + \frac{1}{2}\left[\text{tr}^2M^{(1)} - \text{tr}\,(M^{(1)})^2 \right] + \ldots \right)~,
\end{equation}
while the determinant of $N$ is left unchanged.  The cancellation already established for $T=0$ then gives the finite-temperature continuous sector partition function
\begin{equation}
	Z_\text{cont} = 1 + \text{tr}\,M^{(1)} + \frac{1}{2}\left[\text{tr}^2M^{(1)} - \text{tr}\,(M^{(1)})^2 \right] + \ldots~.
\label{eq:pert_det}
\end{equation}
The matrix $M^{(1)}$ given in (\ref{eq:m1}) depends linearly on the dilaton profile through $\mathcal{L}$ and $\mathcal{R}$ defined in (\ref{eq:mlr}).  These functions are somewhat delicate because the dilaton profile diverges at the boundary.  However, we can get a handle on them by using the connection to the conformal isometry algebra of the Jackiw-Teitelboim model discussed in subsection~\ref{sec:jt:euclid}.  Specifically, they can be recast as matrix elements of the operators $L_0, R_0$ in the $SL(2,\mathbb{R}) \times SL(2,\mathbb{R})$ algebra defined in (\ref{eq:sl2sl2}):
\begin{equation}\begin{aligned}
	\mathcal{L}_{p m,p' m'} &\equiv \int d^2x\,\sqrt{g}\,\epsilon^{\mu\nu}\left(\nabla_\nu \Phi\right) u_{pm} \nabla_\mu u_{p'm'} = \sqrt{\mu a}\langle u_{p m}| L_0 | u_{p' m'}\rangle~, \\
	\mathcal{R}_{p m,p' m'} &\equiv \int d^2x\,\sqrt{g}\,\left(\nabla^\mu \Phi\right) u_{pm} \nabla_\mu u_{p'm'} = \sqrt{\mu a}\langle u_{p m}| R_0 | u_{p' m'}\rangle~.
\label{eq:matrixelems}
\end{aligned}\end{equation}
The leading temperature-dependent piece in the continuous mode partition function (\ref{eq:pert_det}) is then given by
\begin{equation}
	\text{tr}\, M^{(1)} = -\sqrt{\mu a} \sum_m \int dp\, \langle u_{pm} | R_0 | u_{pm}\rangle~.
\label{eq:trm1}
\end{equation}

This matrix element can be evaluated directly via an integral of complex hypergeometric functions over the entire divergent volume of the hyperbolic plane.  However, such an evaluation is in general difficult and requires a careful consideration of how to go to the disk boundary.  We will not do such a computation; instead, there is a simple way to see that the expression (\ref{eq:trm1}) must vanish.  First, recall that (as established in (\ref{eq:rep})) the continuous mode wavefunctions fall into representations of the $SL(2,\mathbb{R})$ such that
\begin{equation}
	\mathbf{L}^2 | u_{pm}\rangle = - \left(p^2 + \frac{1}{4}\right) |u_{pm}\rangle~.
\end{equation}
Then, we can see from the $SL(2,\mathbb{R}) \times SL(2,\mathbb{R})$ algebra (\ref{eq:sl2sl2}) that the operator $R_0$ does not commute with the operator $\mathbf{L}^2$, as defined in (\ref{eq:l2r2}).  Since $|u_{pm}\rangle$ is an eigenfunction of $\mathbf{L}^2$ with an eigenvalue $-(p^2 + \frac{1}{4})$, this implies that $R_0$ cannot preserve the quantum number $p$ when acting on the states $|u_{pm}\rangle$.  That is, $R_0$ must take $|u_{pm}\rangle$ into a combination of other states with \emph{different} values of $p$.  Then, orthogonality of the wavefunctions immediately gives
\begin{equation}
	\langle u_{pm} | R_0 | u_{pm}\rangle = 0~.
\label{eq:r0vanish}
\end{equation}
We can reach the same conclusion via analytic continuation of spherical harmonic matrix elements, as discussed in appendix~\ref{app:cont}.  Either way, we find that
\begin{equation}
	\text{tr}\,M^{(1)} = 0~.
\end{equation}

Summarizing this subsection, we  have $Z_\text{cont} = 1 + \mathcal{O}(\mu a)$, where the higher-order terms require a more complete evaluation of the dilaton matrix elements.  Thus two nice cancellations have occured; namely, the zero-temperature pieces from the physical fields and the ghosts cancel exactly, and the $\mathcal{O}(\sqrt{\mu a})$ terms from the physical fields also cancel. The parameters $\mu$ and $a$ are related to the background black hole quantities such that $\mu a \propto T^2$ so
we are left with
\begin{equation}
	\log Z_\text{cont} = \mathcal{O}(T^2)~.
\end{equation}
In particular, this means that the continuous modes contribute no logarithmic terms to the one-loop free energy. Furthermore, they 
cannot affect the leading-order $\mathcal{O}(T)$ contributions to the black hole entropy from thermal fluctuations.

\subsection{Discrete Modes}
\label{sec:oneloop:harm}

In this subsection we tackle the contribution of the discrete modes to the one-loop partition function. These modes are orthogonal to the continuous modes considered in the previous subsection and this decoupling persists when a dilaton profile which, as we argued there (and in appendix~\ref{app:cont}), is equivalent to a small change in the continuous quantum number.  We therefore focus on the discrete modes by themselves, arising from quantum inequivalence when dualizing either vector or tensor fields entirely to scalars. There are two types: harmonic modes (vectorial) and quadratic holomorphic differential forms (tensorial). We consider them in turn. 

The vector harmonic modes are 
\begin{equation}
	\mathcal{A}_z^{(n)} = \ell\sqrt{\frac{n}{2\pi}} z^{n-1}~,
\end{equation}
plus the Hermitian conjugates, where $n= 1, 2,\ldots$, and the normalization has been chosen such that $\langle \mathcal{A}^{(n)} | \mathcal{A}^{(m)}\rangle = \ell^2 \delta_{n,m}$.  The configuration space of the ghost sector includes these discrete modes because the ghosts are vector fields. However, we chose a gauge in which the ghosts do not couple to the background dilaton. Therefore, the ghost fields do not contribute to the logarithmic dependence of the one-loop partition function on the temperature and so we do not need to consider them in detail for our purposes. 

It is also important to take the harmonic modes into account when considering the configuration space of the traceless symmetric tensor. They give rise to the tensor modes
\begin{equation}\begin{aligned}
	\mathcal{B}^{(n)}_{zz} &= \ell\nabla_z \mathcal{A}^{(n)}_z = \ell^2\sqrt{\frac{n}{2\pi}} z^{n-2} \frac{n(1-|z|^2)-1-|z|^2}{1-|z|^2}~,
	\label{eqn:Bzztensor}
\end{aligned}\end{equation}
plus their Hermitian conjugates. These modes are ``pure gauge" by construction because they arise from the diffeomorphism vector $\mathcal{A}_z^{(n)}$. Moreover, $\mathcal{A}_z^{(n)}$ is not a large diffeomorphism because it is normalizable. In simpler circumstances these field configurations would be discarded as unphysical, or they would be cancelled by ghosts. However, they couple non-trivially to the dilaton background, and so (unlike the ghosts discussed in the previous paragraph) we must retain these discrete modes and analyze them in detail. 

The quadratic holomorphic differentials (QHDs) also give discrete tensor modes
\begin{equation}
	\mathcal{H}^{(n)}_{zz} = \ell^2\sqrt{\frac{n(n^2-1)}{2\pi}}\,z^{n-2}~,
\end{equation}
with $n = 2,3,\ldots$.  These QHD modes and the tensor modes \eqref{eqn:Bzztensor} constructed from harmonic vectors together form an orthonormal basis 
\begin{equation}
	\langle \mathcal{B}^{(n)} | \mathcal{B}^{(m)}\rangle = \ell^2 \delta_{n,m}~, \quad \langle \mathcal{H}^{(n)} | \mathcal{H}^{(m)} \rangle = \ell^2 \delta_{n,m}~, \quad \langle \mathcal{B}^{(n)} | \mathcal{H}^{(m)}\rangle = 0~,
	\label{eqn:orthonormal}
\end{equation}
for the discrete tensor modes. The discrete part of the traceless, symmetric part of the graviton $\bar{h}_{\mu\nu}$ can therefore be expanded as
\begin{equation}
	\bar{h}_{zz} = \sum_{n=1}^\infty b_{1 n} \mathcal{B}_{zz}^{(n)} + \sum_{n=2}^\infty b_{2n} \mathcal{H}^{(n)}_{zz}~,
\label{eq:hdecomp}
\end{equation}
for some arbitrary complex constants $b_{i n}$.

Our next step is to insert this expansion into the (Euclideanized) quadratic action (\ref{eq:squad}). The orthonormality relations \eqref{eqn:orthonormal} between discrete tensor modes decouple the harmonic modes and the QHDs on pure AdS$_2$ . Conveniently, this decoupling remains even after the dilaton profile is taken into account because integrals of the form
\begin{equation}
	\int dz\,d\bar{z}\,\sqrt{g}\,\left(\Phi - 1\right) \mathcal{H}^{(n)}_{zz} \mathcal{B}^{(m)z z}~, \quad \int dz\,d\bar{z}\,\sqrt{g}\,\nabla^\mu \Phi \nabla_\mu \mathcal{H}^{(n)}_{zz} \mathcal{B}^{(m)z z}~,
\end{equation}
vanish. This means we can split up the one-loop quadratic action (\ref{eq:squad}) as
\begin{equation}
	I = I_\text{harmonic} + I_\text{QHD}~.
\end{equation}
After explicitly inserting the discrete mode decomposition (\ref{eq:hdecomp}) into the quadratic action we now find 
\begin{equation}\begin{aligned}
	I_\text{harmonic} &= \sum_{n=1}^\infty |b_{1 n}|^2\left(\frac{1}{8\pi G_2 \xi}-\frac{5n\sqrt{\mu a}}{16\pi G_2}\right)~, \\
	I_\text{QHD} &= \sum_{n=2}^\infty \frac{|b_{2n}|^2 n \sqrt{\mu a}}{16\pi G_2}~.
	\label{eqn:harmQHD}
\end{aligned}\end{equation}
Note that the term due to the QHDs vanishes in the limit of vanishing dilaton slope $\sqrt{\mu a} \to 0$. This is because these modes are true zero-modes from the perspective of Euclidean AdS$_2$.  

The one-loop partition functions are now easily computed as Gaussian integrals over the expansion coefficients $b_{1n}$ and $b_{2n}$:
\begin{equation}\begin{aligned}
	Z_\text{harmonic} &= \prod_{n=1}^\infty \int d b_{1n} d \bar{b}_{1n} \,\text{exp}\left(-|b_{1n}|^2 \frac{2-5 n \xi \sqrt{\mu a}}{16\pi G_2\xi}\right) = \prod_{n=1}^\infty \frac{32\pi^2 G_2 \xi}{2- 5n \xi \sqrt{\mu a}}~, \\
	Z_\text{QHD} &= \prod_{n=2}^\infty \int d b_{2n}\, d\bar{b}_{2n}\,\text{exp}\left(-|b_{2n}|^2 \frac{n \sqrt{\mu a}}{16\pi G_2}\right) = \prod_{n=2}^\infty \frac{32 \pi^2 G_2}{n \sqrt{\mu a}}~.
\end{aligned}\end{equation}
Using our classical dictionary (\ref{eq:bhmass}) to relate the parameters $\mu,a$ to the physical scales in the theory we then find 
\begin{equation}\begin{aligned}
	\log Z_\text{harmonic} &= -\sum_{n=1}^\infty \log \left(\frac{1}{16\pi^2 G_2 \xi} -\frac{5 n T}{8 \pi^2 M_\text{gap}}\right) ~, \\
	\log Z_\text{QHD} &= - \sum_{n=2}^\infty \log \left(\frac{n T}{8\pi^2 M_\text{gap}}\right)~.
\label{eq:zdisc}
\end{aligned}\end{equation}

As is typical in one-loop computations, our results for the partition functions are divergent so we must carefully regulate these expressions and extract the finite, physical terms.  It is only the logarithmic term in the free energy that come entirely from one-loop fluctuations, and this is what we want to compute.  Moreover, we focus on the temperature-dependence of this logarithmic piece in order to make connection with the SYK model.  One can argue by dimensional analysis that the regularized one-loop partition function must take the form
\begin{equation}
	\log Z = \alpha \log \frac{T}{\Lambda} + c_0 + c_1 \frac{T}{\Lambda} + c_2 \frac{T^2}{\Lambda^2} + \ldots~,
\label{eq:zexp}
\end{equation}
where $\Lambda$ is some renormalization scale and $\alpha, c_i$ are some Wilsonian numerical parameters. The $\mathcal{O}(1)$ terms are scheme-dependent and depend on precisely how we regularize our one-loop divergences. However, the coefficient $\alpha$ is unambiguous and can be extracted by computing
\begin{equation}
	\lim_{T\to 0} T \frac{\partial \log Z}{\partial T} = \alpha~.
\end{equation}
We can use this relation to extract the coefficient of the logarithmic term for the harmonic and QHD contributions to the partition function 
(\ref{eq:zdisc}).  The result of this is
\begin{equation}\begin{aligned}
	&\lim_{T\to 0} T \frac{\partial \log Z_\text{harmonic}}{\partial T} = 0~, \\
	&\lim_{T\to 0} T \frac{\partial \log Z_\text{QHD}}{\partial T} = - \sum_{n=2}^\infty 1 = 1 - \zeta(0) \cong \frac{3}{2}~,
	\label{eqn:finalcomp}
\end{aligned}\end{equation}
where in the last line we used zeta function regularization to make the sum over QHD modes finite. The technical reason that the harmonic modes do not contribute is that the temperature $T\sim \sqrt{\mu a}$ appears as a shift of the nonvanishing eigenvalues in \eqref{eqn:harmQHD}, in contrast to the QHDs that acquire their entire ``mass" from the breaking of conformal symmetry. 
This is a precise version of the physical reasoning based on effective quantum field theory that was advanced in subsection \ref{sec:EFTE} and especially in figure~\ref{fig:scales2}.

Therefore, we find that the logarithmic corrections to the partition function arise entirely from the quadratic holomorphic differentials:
\begin{equation}
	\log Z_\text{QHD} = \frac{3}{2}\log \frac{T}{\Lambda} + \ldots~,
\end{equation}
where the dots indicate terms that are independent of $T$ or subleading in $T$.  Since the size of the dilaton profile (and thus the conformal symmetry breaking scale) is controlled by the parameter $\sqrt{\mu a} = 4 G_2 T / M_\text{gap}$ we identify the renormalization scale $\Lambda$ as
\begin{equation}
	\Lambda = \frac{M_\text{gap}}{G_2}~.
\end{equation}
We have required $G_2 \ll 1$ (for many reasons, including in \eqref{eqn:hier2}) and thus $\Lambda\gg M_\text{gap}$. This condition allows a regime of temperatures 
$\Lambda\gg T\gg M_\text{gap}$ where it is justified to treat the Jackiw-Teitelboim model as an effective field theory, as indicated in \eqref{eqn:hier2}.

The QHDs determine the entire logarithmic term in our partition function since the continuous modes and the harmonic modes do not contribute.  Therefore, our final result for the one-loop partition function of the Jackiw-Teitelboim model is
\begin{equation}
	\boxed{\left. \log Z \right|_{\rm one-loop}= \frac{3}{2} \log \frac{G_2 T}{M_\text{gap}} = - \frac{3}{2} \log \frac{\beta M_\text{gap}}{G_2}}~.
\label{eq:result}
\end{equation}
This one-loop partition function, computed entirely using bulk methods, is the main result of our paper.  It matches the one-loop SYK partition function (\ref{eq:syk_z}) and serves as a precision test for the near-AdS$_2$/near-CFT$_1$ correspondence.  

\section{Black Hole Entropy}
\label{sec:bhentropy}

Having the one-loop correction to the partition function, we can determine the corresponding one-loop correction to the entropy of black holes in the Jackiw-Teitelboim model.  The partition function is computed in the canonical ensemble, where the temperature of the system is fixed, while the entropy is computed in the microcanonical ensemble, where each state has a fixed energy.  In this section, we will perform the Legendre transform from the canonical ensemble to the microcanonical ensemble in order to determine the one-loop correction to the Bekenstein-Hawking law (\ref{eq:bhlaw}).  This result has been derived before~\cite{Maldacena:2016hyu}, but we find it instructive to go through the details and discuss implications of the result.

Let $\mathcal{H}$ denote the Hilbert space of black hole microstates $|i\rangle$, each with an associated energy $E_i$.  The canonical partition function can then be presented as
\begin{equation}
	Z(\beta) = \sum_{i \in \mathcal{H}} \langle i | e^{-\beta E_i} | i\rangle~.
\label{eq:grandtrace}
\end{equation}
When the black hole is large, the interactions between the black hole and any fields in the system are negligible compared to the black hole's mass energy, and so we can approximate each microstate's energy as $E_i \approx M$.  The canonical partition function can then be expressed as a sum over all black hole microstates, weighted by the microcanonical degeneracy of microstates $\Omega(M) = e^{S(M)}$, such that
\begin{equation}
	Z(\beta) = \int dM\, e^{S(M) - \beta M}~.
\label{eq:zsum}
\end{equation}
The integrand in (\ref{eq:zsum}) is strongly peaked around the classical value of $M$ corresponding to a given inverse temperature $\beta$.  We can therefore perform a Laplace transformation to invert this expression and solve for the microcanonical entropy, which yields
\begin{equation}
	e^{S(M)} = \int d\beta\, Z(\beta)\, e^{\beta M}~.
\label{eq:smc}
\end{equation}

We consider sufficiently large black holes such that the logarithmic quantum correction to the entropy dominates over all other corrections. Then the saddle-point approximation is justified when computing the integral in (\ref{eq:smc}). It gives
\begin{equation}
	e^{S(M)} \approx Z(\beta_\text{cl})e^{\beta_\text{cl}M} \sqrt{2\pi}\left(\frac{\partial ^2 \log Z(\beta_\text{cl})}{\partial \beta ^2}\right)^{-1/2}~,
\end{equation}
where $\beta_\text{cl}$ is the classical value of the inverse temperature $\beta$ that corresponds to a black hole of mass $M$. Inserting partition function $\log Z(\beta) = - I_\text{cl} + \log Z(\beta)\big{|}_\text{one-loop}$ including the one-loop correction \eqref{eq:result} we find that
\begin{equation}
	S(M) \approx -I_\text{cl} + \beta_\text{cl} M + \log Z(\beta_\text{cl})\big{|}_\text{one-loop} - \frac{1}{2}\log\left(\frac{\partial ^2 \log Z(\beta_\text{cl})}{\partial \beta ^2}\right) + \mathcal{O}(1)~.
\label{eq:legendre}
\end{equation}
The first two terms in this expression are the classical contributions which combine to the Bekenstein-Hawking entropy $S_\text{BH}$. The next two terms are the logarithmic quantum corrections.  
The classical partition function $\log Z(\beta_\text{cl})$ is linear in the temperature $T=\beta^{-1}$ so
\begin{equation}
	\frac{\partial ^2 \log Z(\beta_\text{cl})}{\partial \beta ^2} \sim \beta^{-3}~.
\end{equation}
However, the one-loop partition function $\log Z(\beta_\text{cl})\big{|}_\text{one-loop} \sim -\frac{3}{2}\log \beta$ so we find that the logarithmic contributions to the entropy coming from one-loop corrections to the free energy (\ref{eq:result}) {\it cancel} precisely with the logarithmic terms we obtain from performing the Legendre transform (\ref{eq:legendre}). We are simply left with
\begin{equation}
	S(M) = S_\text{BH} = \frac{\Phi|_H}{4 G_2}~,
\label{eq:slog}
\end{equation}
with {\it no} logarithmic corrections to the microcanonical entropy.

This cancellation is somewhat surprising.  The coefficient of the logarithmic correction sets a dynamical scale in quantum gravity that runs along the RG flow~\cite{PERRY1978114,Tong:2014era,Charles:2017dbr}.  When the background preserves supersymmetry, this coefficient can be related via index theorems to certain topological invariants on the background manifold and thus it becomes protected along the RG flow~\cite{Christensen:1978md,Vassilevich:2003xt,Banerjee:2010qc,Sen:2014aja}.  Recent results show that there are situations in which this coefficient is topological even on non-BPS backgrounds~\cite{Charles:2015eha,Castro:2018hsc}, but this non-renormalization still relies on supersymmetry at the level of the action itself~\cite{Charles:2017dbr}.  The Jackiw-Teitelboim model is not supersymmetric, though, yet it appears that the logarithmic corrections arising from the gravitational sector of the near-horizon geometry are protected. This merits further exploration.

\section{Discussion}
\label{sec:conc}

In this paper we have computed the logarithmic temperature dependence $-\frac{3}{2}\log \beta$ in the partition function of the JT model. 
This is a test of near-AdS$_2$/near-CFT$_1$ holography because it agrees precisely with the one found in the SYK model. 

Our result is far from new at the purely numerical level, as the same answer has been reported many times over the last few years. However, previous computations exploited the classical equivalence of the JT model with the Schwarzian boundary theory at low temperature and then analyzed quantum corrections in the latter theory. In contrast, we use traditional methods in Euclidean quantum gravity. Our different perspective highlights several points that were not emphasized in recent literature and would be worth developing further.

\paragraph{Quantum corrections to extremal black hole entropy.}

The full logarithmic term in the partition function (\ref{eq:result}) can be written as
$$
\log Z |_{\text{one-loop}} = \frac{3}{2}\log\left( \frac{T}{M_{\rm gap} S_0}\right)~. 
$$
The entropy above extremality $\Delta S = S-S_0 = TM^{-1}_{\rm gap}$ is the classical entropy described by the Schwarzian at finite temperature, and quantum fluctuations around this saddle point give the correct temperature dependence $\frac{3}{2}\log (TM^{-1}_{\rm gap})$. However, the additional term $-\frac{3}{2}\log S_0$ that we find can not be extracted from the Schwarzian because it involves an entirely different dimensionless parameter. 

This additional contribution $-\frac{3}{2}\log S_0$ to the partition function is important for the precision comparison between microscopic and macroscopic entropy for BPS black holes with AdS$_2\times S^2$ near-horizon geometry. For such extremal black holes $T = 0$ and so the Schwarzian modes must be integrated out as we lower the effective cut-off scale $\Lambda$ below $M_{\rm gap}$.  The $-\frac{3}{2}\log S_0$ term is the threshold correction from integrating out these modes. It joins contributions of the form $\#\log S_0$ that come from Kaluza-Klein modes (including Killing vectors on the $S^2$) that were integrated out already when lowering the effective scale $\Lambda$ below $\Lambda_\text{KK} \sim \ell^{-1}$ and, together, they give the logarithmic quantum correction to the ground state entropy~\cite{Sen:2014aja,Keeler:2014bra}. 

Thus our computation clarifies the connection between recent SYK and JT results to previous work on quantum corrections to extremal black hole entropy. 

\paragraph{Quadratic holomorphic differentials.}  According to our method, the only modes that ultimately contribute to the logarithmic term in the one-loop partition function are the QHDs $\mathcal{H}^{(n)}_{\mu\nu}$, where $n = 2,3,\ldots$.  These are exact zero-energy modes in the true infrared that acquire a small mass by finite temperature effects, which in turn yields a large logarithm in the partition function. The analogous computation in the SYK model organizes field configurations by their weight $h$ under the conformal Casimir under which the modes with weight $h=2$ have vanishing eigenvalue at zero temperature, but finite temperature effects shift it slightly above zero. These modes, and their relatives in the Schwarzian, are also enumerated by $n= 2,3,\ldots$ and sum up to give the $-\frac{3}{2}\log \beta$ term in the partition function, just like the QHDs in bulk.  

Importantly, despite the obvious similarity between these computations, the modes that contribute are not the same; they are instead \emph{dual} to one another. The reparameterization symmetry that is emergent in the SYK model and described by the Schwarzian is ${\rm Diff}(S^1)/SL(2)$ which is generated by smooth vector fields on $S^1$. The adjoints of these vector fields are precisely the quadratic holomorphic differentials and should be thought of as smooth tensors on $S^1$ that describe deformations of the bulk geometry~\cite{Witten:1987ty,Taylor:1992xt}. This duality is akin to the relation between the loop-space approach to string theory and the Polyakov path integral. 

Our perspective may be particularly appropriate for the recent and exciting developments relating random matrix theory on disconnected boundaries to baby universes with complicated topology in bulk~\cite{Saad:2019lba,Stanford:2019vob}. There, the perturbative contributions to the bulk theory involve integrating over the Weil-Petersson measure on the Teichm\"{u}ller space of Riemann surfaces with any genus. The all-important novelty is the coupling to boundaries via the introduction of ``trumpets" endowed with a Schwarzian theory on their outer boundary. Representing the space of trumpet deformations in terms of quadratic holomorphic differentials, as we do, puts the outer boundary on equal footing with the handles and the inner boundaries due to D-branes. This geometric interpretation of the entire gravitational theory, including the trumpets, may confer some advantages. It would be interesting to test this construction by computing correlation functions in this formalism.

\paragraph{Quantum universality.}  Many authors have argued that the JT model captures universal aspects of near-extremal black holes but most evidence we are aware of involves classical field theory, and in particular the classical equivalence of various 2D dilaton gravity theories. Our computation does not use special features of JT gravity and after quantization the couplings to the background are quite complicated, as one expects generically. Nonetheless, the logarithmic correction $-\frac{3}{2}\log \beta$ from the dilaton gravity sector will generalize to all other near-extreme black holes. 

This expectation is based on effective quantum field theory and illustrated in figure \ref{fig:scales2}.  For gravitational theories on AdS$_2$, couplings to the background curvature introduce a mass shift of $2 \ell^{-2}$ for the quadratic holomorphic differential modes.  We showed this explicitly in subsection~\ref{sec:quadact}, but this has also been noted in previous work~\cite{Keeler:2014bra,Castro:2018ffi,Castro:2019crn}.  This shifts the effective mass of the QHDs to be precisely zero.  Breitenlohner-Freedman's stability criterion requires non-negative effective mass, and so the quadratic holomorphic differentials precisely saturate this bound.  The QHD contribution to the partition function will therefore always dominate at sufficiently low energy. Moreover, their number is topological, because it is counted by the Riemann-Roch theorem. Their contribution $-\frac{3}{2}\log \beta$ to the partition function is thus universal for any near-AdS$_2$ theory of gravity. 

\section*{Acknowledgments}
We thank Robert McNees, Miguel Montero, Alex Streicher, Thomas Van Riet, and Herman Verlinde for useful discussions.  A preliminary version of this work was presented at the TSIMF workshop on Black Holes and Holography in January 2018.  This work was supported by the U.S. Department of Energy under grant DE-SC0007859. AMC is supported in part by the KU Leuven C1 grant ZKD1118 C16/16/005, the National Science Foundation of Belgium (FWO) grant G.001.12 Odysseus, and by the European Research Council grant no. ERC-2013-CoG 616732 HoloQosmos.

\appendix

\section{Other Approaches}
\label{app:z}

In this appendix, we review previous computations of the one-loop partition function, first in the SYK model and then in the one-dimensional Schwarzian theory.

\subsection{The SYK Model Analysis}

The analysis of~\cite{Maldacena:2016hyu} identified the approximate zero modes in the SYK model as those coming from the $h=2$ modes (where $h$ is their weight under the conformal Casimir) and determined their 
thermal mass as $\frac{|n|}{\beta J}$ with $n=\pm 2, \pm 3,\ldots$. The path integral therefore acquires the contribution
\begin{equation}
	\log Z = -\sum_{n=2}^\infty \log\left(\frac{n}{\beta J}\right)~,
\end{equation}
up to constants that are independent of the coupling $\beta J$.  The dependence on $n$ amounts to a contribution that is divergent but independent of $\beta J$, 
and thus does not constitute a physical contribution to the partition function.  We can then extract the $\beta J$-dependence using zeta function regularization, as in previous computations. This gives
\begin{equation}
	\log Z = \log \beta J \left(-1 + \sum_{n=1}^\infty 1\right) = \log \beta J\left(\zeta(0) - 1\right) = -\frac{3}{2}\log \beta J~.
\label{eq:zsyk}
\end{equation}

The final manipulation yielding the quantum corrections in the SYK model is essentially the same that we do in \eqref{eqn:finalcomp}. However, the origin of the $h=2$ modes in the SYK model is reparametrization invariance of time in a quantum mechanical model while for us the starting point is quantum gravity. In contrast, we gauge-fix a path integral over all metrics and ultimately identify the quadratic holomorphic modes as the contributors to the logarithmic quantum correction. 

\subsection{The Schwarzian Analysis}

The finite-temperature Schwarzian action, following the conventions of~\cite{Maldacena:2016hyu}, is
\begin{equation}
	I[\epsilon] = \frac{\alpha_S N}{2 \mathcal{J}} \int_0^{\beta} d\tau\,\bigg{[} (\epsilon'')^2 - \left(\frac{2\pi}{\beta}\right)^2 (\epsilon')^2\bigg{]}~,
\end{equation}
where $\alpha_S$ is a numerical factor related to the spectrum of the SYK model, $N$ is the number of Majorana fermions in the SYK model, and $\mathcal{J} \propto J$.  This action has an $SL(2,\mathbb{R})$ reparameterization symmetry that acts as a gauge symmetry on the allowed $\epsilon$ configurations.  The partition function of the theory is thus given by
\begin{equation}
	Z = \int \frac{\mathcal{D}\epsilon}{\text{Vol}(SL(2,\mathbb{R}))} e^{-I[\epsilon]}~,
\end{equation}
where we have modded out by the volume of the $SL(2,\mathbb{R})$ symmetry in order to have a well-defined gauge orbit for the path integral.  $SL(2,\mathbb{R})$ is a three-dimensional group, and so we can fix the gauge completely by specifying what values $\epsilon$, $\epsilon'$, and $\epsilon''$ take at $\tau = 0$.  We therefore can express the path integral as
\begin{equation}
	Z = \int \mathcal{D}\epsilon\,\delta(\epsilon(0))\delta(\epsilon'(0))\delta(\epsilon''(0)) e^{-I[\epsilon]}~.
\end{equation}
We now define the rescaled fields and parameters
\begin{equation}
	\tilde{\tau} = \frac{\tau}{\beta}~, \quad \tilde{\epsilon}(\tilde{\tau}) = \frac{\epsilon(\tau)}{\beta}~,
\end{equation}
such that $\tilde{\tau}$ and $\tilde{\epsilon}$ are dimensionless.  The action becomes
\begin{equation}
	I[\tilde{\epsilon}] = \frac{\alpha_S N}{2 \beta \mathcal{J}} \int_0^1 d \tilde{\tau}\,\bigg{[}(\tilde{\epsilon}'')^2 - (2\pi \tilde{\epsilon}')^2 \bigg{]}~.
\end{equation}
The product of delta functions in the path integral is invariant under this change of variables, and so the partition function can therefore be rewritten as
\begin{equation}
	Z = \int \mathcal{D} \tilde{\epsilon}\, \delta(\tilde{\epsilon}(0))\delta(\tilde{\epsilon}'(0))\delta(\tilde{\epsilon}''(0)) e^{-I[\tilde{\epsilon}]}~.
\end{equation}
The strategy now is to rescale the fields such that the action has no dependence on $\beta \mathcal{J}$.  This will ensure that all of the $\beta\mathcal{J}$-dependence will appear in the gauge-fixing part of the path integral, which we can then easily extract.  If we define the field
\begin{equation}
	\phi(\tilde{\tau}) = \frac{\tilde{\epsilon}(\tilde{\tau})}{\sqrt{\beta \mathcal{J}}}~,
\end{equation}
then the action for $\phi$ is entirely independent of $\beta \mathcal{J}$ and the product of delta functions now becomes
\begin{equation}
	\delta(\tilde{\epsilon}(0))\delta(\tilde{\epsilon}'(0))\delta(\tilde{\epsilon}''(0)) = (\beta \mathcal{J})^{-3/2}\delta(\phi(0))\delta(\phi'(0))\delta(\phi''(0))~.
\end{equation}
The partition function is then given by
\begin{equation}
	Z = (\beta \mathcal{J})^{-3/2}\int \mathcal{D}\phi\,\delta(\phi(0))\delta(\phi'(0))\delta(\phi''(0)) e^{-I[\phi]}~,
\end{equation}
where the path integral is now independent of $\beta \mathcal{J}$.  And, since $J \propto \mathcal{J}$, we can take the logarithm of both sides to find that
\begin{equation}
	\log Z = -\frac{3}{2}\log \beta J~,
\end{equation}
up to constant terms that are independent of $\beta J$.  

Interestingly, the $-3$ in the final answer is due, in this computation, to $3$ ``missing" configurations that would be equivalent under $SL(2,\mathbb{R})$.  In contrast, our approach focuses on the ``present" configurations that are due to large diffeomorphisms realized by the quadratic holomorphic differentials. They are infinite in number in a manner that renormalizes to $-3$ net configurations. 

The Schwarzian is a one-loop exact theory, so this result captures the quantum corrections to the partition function to all orders.  However, this does not mean that the JT model and the SYK model are one-loop exact; rather, the Schwarzian is meant to capture the low-temperature dynamics of these theories that arise from their conformal symmetry breaking patterns.

\section{Analytic Continuation from \texorpdfstring{$S^2$}{S2} to \texorpdfstring{$H^2$}{H2}}
\label{app:cont}

In this appendix, we analytically continue results from the sphere $S^2$ onto the hyperbolic space $H^2$ in order to leverage elementary results from the study of the rotation group and illustrate conceptual and practical aspects of $SL(2)$ symmetry, including conformal symmetry, continuous mode wavefunctions, and dilaton matrix elements.

Consider the line element on $S^2$:
\begin{equation}
	ds^2_{S^2} = \ell^2\left(d\theta^2 + \sin^2\theta\,d\phi^2\right)~.
\end{equation}
Under the identification $\eta = i \theta$, this metric becomes the negative of the line element for two-dimensional hyperbolic space $H^2$:
\begin{equation}
	ds^2_{H^2} = \ell^2\left(d\eta^2 + \sinh^2\eta\,d\phi^2\right)~.
\end{equation}
The metric on hyperbolic space can be further transformed to the disk metric in holomorphic coordinates (\ref{eq:disk_metric}) by the change of variables $z = \tanh\left(\eta/2\right) e^{i\phi}$.  Moreover, we can choose coordinates such that the geometry for the black holes in the Jackiw-Teitelboim model takes precisely this form. This indicates that there must be a way to analytically continue quantities on $S^2$ onto related quantities on the Jackiw-Teitelboim background (\ref{eq:disk_metric}).

As an example, we can generate the background dilaton profiles in the Jackiw-Teitelboim model by this procedure.  The starting point is to place the action (\ref{eq:sjt}) on $S^2$ by flipping the sign of the cosmological constant term.  The corresponding equations of motion are then
\begin{equation}
	R = \frac{2}{\ell^2}~, \quad \nabla_\mu \nabla_\nu \Phi = -\frac{g_{\mu\nu}}{\ell^2} (\Phi - 1)~.
\end{equation}
The dilaton equation of motion can be integrated exactly to yield solutions $\Phi = 1 + Y$ for some profiles $Y$.  
There are precisely three linearly independent solutions which can be identified with the usual spherical harmonics $Y^1_m$ where $m=-1, 0, 1$. We choose the
basis 
\begin{equation}
	Y_{-1} = i \sin\theta\,e^{i\phi}~, \quad Y_0 = i \cos\theta~, \quad Y_{+1} = i \sin\theta\,e^{-i\phi}~.
\end{equation}
If we perform the analytic continuation $\eta = i \theta$, go to holomorphic coordinates, and then compare the results to the Jackiw-Teitelboim dilaton profiles (\ref{eq:xprofiles}), we find the identifications
\begin{equation}\begin{aligned}
	Y_{-1} &\rightarrow X_{-1}~, \\
	Y_{0} &\rightarrow i X_0~, \\
	Y_{+1} &\rightarrow X_{+1}~.
\label{eq:yxmap}
\end{aligned}\end{equation}
Thus analytic continuation gives a precise map between the dilaton profiles in the Jackiw-Teitelboim model and familiar functions on $S^2$.

In subsection~\ref{sec:jt:euclid} we showed that the dilaton profiles in the Jackiw-Teitelboim model are central to understanding the symmetry of the theory
because they generate Killing vectors $\zeta^\mu_n$ and conformal Killing vectors $\xi^\mu_n$ on $H^2$.  The Killing vectors form the $SL(2,\mathbb{R})$ isometry algebra of the background $H^2$ metric which can be extended to $SL(2,\mathbb{R}) \times SL(2,\mathbb{R})$ by adding the conformal Killing vectors.  

In the analogous procedure on $S^2$ we first define the vector fields:
\begin{equation}\begin{aligned}
	\zeta^\mu_z &= \epsilon^{\mu\nu} \nabla_\nu Y_0~, &&\quad \zeta^\mu_\pm &&= \epsilon^{\mu\nu} \nabla_\nu Y_\pm~,\\
	\xi^\mu_z &= \nabla^\mu Y_0~, &&\quad \xi^\mu_{\pm} &&= \nabla^\mu Y_{\pm}~.
\end{aligned}\end{equation}
The $\zeta^\mu_n$ are the three Killing vectors on $S^2$ while the $\xi^\mu_n$ are the three globally defined conformal Killing vectors on $S^2$.
The Lie derivatives with respect to these vectors define the operators
\begin{equation}
	\begin{split}\begin{aligned}[t]
		L_z &= \ell^2\mathcal{L}_{\zeta_0} = \ell^2(\epsilon^{\mu\nu} \nabla_\nu Y_0) \nabla_\mu~,\quad\\
		R_z &= \ell^2\mathcal{L}_{\xi_0} = \ell^2(\nabla^\mu Y_0) \nabla_\mu~,\quad 		
	\end{aligned}
	\begin{aligned}[t]
			L_\pm &= \ell^2\mathcal{L}_{\zeta_\pm} = \ell^2(\epsilon^{\mu\nu}\nabla_\nu Y_\pm)\nabla_\mu~,\\
			R_\pm &= \ell^2\mathcal{L}_{\xi_\pm} = \ell^2(\nabla^\mu Y_\pm)\nabla_\mu~.
	\end{aligned}\end{split}
\end{equation}
Their explicit forms are 
\begin{equation}
	\begin{split}\begin{aligned}[t]
		L_{+} &= e^{-i\phi}\left(\partial_\theta - i \cot\theta\,\partial_\phi\right)~, \quad \\
		L_{z} &= i \partial_\phi~, \quad \\
		L_{-} &= e^{i\phi}\left(-\partial_\theta - i \cot\theta\,\partial_\phi\right)~, \quad \\
	\end{aligned}
	\begin{aligned}[t]
		R_{+} &= e^{-i\phi}\left(i \cos\theta\,\partial_\theta + i \csc\theta\,\partial_\phi\right)~, \\
	R_{z} &= - i\sin\theta\, \partial_\theta~, \\
	R_{-} &= e^{i\phi}\left(i \cos\theta\,\partial_\theta - \csc\theta\,\partial_\phi\right)~.
	\end{aligned}\end{split}
\end{equation}
These operators satisfy the commutation relations:
\begin{equation}
	\begin{split}\begin{aligned}[t]
		[L_z, L_{\pm}] &= \pm L_{\pm}~, \quad \\
		[L_z, R_{\pm}] &= \pm R_{\pm}~, \quad \\
		[R_z, L_{\pm}] &= \pm R_{\pm}~, \quad \\
		[R_z, R_{\pm}] &= \pm L_{\pm}~, \quad 
	\end{aligned}
	\begin{aligned}[t]
		[L_{+}, L_{-}]&= 2 L_z~, \\
			[L_{+}, R_{-}]&= 2 R_z~, \\
			[R_{+}, L_{-}]&= 2 R_z~,\\
			[R_{+}, R_{-}]&= 2 L_z~.
	\end{aligned}\end{split}
\label{eq:su2su2}
\end{equation}
The first line identifies the operators $L_n$ originating from the Killing vectors as the familiar $SU(2)$ algebra which expresses rotational symmetry of the background $S^2$ metric. The next two lines show that the operators $R_n$ due to the conformal Killing vectors are in fact vectors under the rotation group, as expected. 
The final line closes the algebra in a manner such that, as a whole, it can be recast as two copies of $SU(2)$. This can be seen by forming the linear combinations $J^\pm_n = \frac{1}{2}\left(L_n \pm R_n\right)$ and checking that the $J^+_n$ and $J^-_n$ operators each obey $SU(2)$ commutation relations and commute with each other. 

This enhancement of symmetry has an instructive analogue in classical mechanics.  In three dimensions, when an object is subject to a central force, the system has rotational symmetry and the angular momentum $\mathbf{L}$ is conserved.  When the force is the scale free inverse-square force $\mathbf{F} = - k \mathbf{r} / r^3$, we can define the Laplace-Runge-Lenz vector~\cite{doi:10.2991/jnmp.2003.10.3.6} 
\begin{equation}
	\mathbf{R} = \mathbf{p} \times \mathbf{L} - \frac{m k \mathbf{r}}{r}~,
\end{equation}
where $\mathbf{p}$ is the momentum, $\mathbf{L}$ is the angular momentum, and $m$ is the mass of the object.  This vector is important because, along with the energy and angular momentum of the object, $\mathbf{R}$ is also conserved.  When we account for a constraint relating these six conserved quantities we find five constants of motion, which is precisely the amount required to completely determine a trajectory in three dimensions.

Upon quantization, the components of the angular momentum $\mathbf{L}$ become operators that form an $SU(2)$ algebra.  When we additionally quantize the components of the Laplace-Runge-Lenz vector $\mathbf{R}$, though, the full algebra is $SO(4) \cong SU(2) \times SU(2)$~\cite{refId0}; this is precisely the algebra shown in~(\ref{eq:su2su2}).  This enhancement of symmetry is well-known for any three-dimensional quantum-mechanical system with a $1/r$ potential in the Hamiltonian, and is commonly referred to as the ``hidden" $SO(4)$ symmetry of the Hydrogen atom. It is the reason that all states with the same principal quantum number $n=1,2,\ldots$ have the same energy, independent of the value of the angular momentum $l=0,\ldots, n-1$. 

The Laplacian on $S^2$ is related to the operators $\mathbf{L}^2$, $\mathbf{R}^2$, defined as follows:
\begin{equation}\begin{aligned}
	\ell^2 \square &= -\mathbf{L}^2 \equiv - L_z^2 - \frac{1}{2}\left(L_{-} L_{+} + L_{+} L_{-}\right) \\
	&= -\mathbf{R}^2 \equiv -R_z^2 - \frac{1}{2}\left(R_{-} R_{+} + R_{+} R_{-}\right)~.
\end{aligned}\end{equation}
The corresponding eigenfunctions of this Laplacian are the spherical harmonics $Y_l^m$, which in our conventions are given by
\begin{equation}
	Y_l^m(\theta,\phi) = N_{lm} P_l^m(\cos\theta) e^{im\phi}~, \quad N_{lm} = \sqrt{\frac{(2l+1)}{4\pi} \frac{(l-m)!}{(l+m)!}}~.
\end{equation}
Specifically, these are eigenfunctions of the operators $\mathbf{L}^2$ and $L_z$:
\begin{equation}\begin{aligned}
	\mathbf{L}^2 |Y_l^m\rangle &= l(l+1)|Y_l^m\rangle~, \\
	L_z |Y_l^m\rangle &= -m |Y_l^m\rangle~.
\end{aligned}\end{equation}
As shown in~\cite{Camporesi:1994ga}, we can continue these states into eigenfunctions $|u_{pm}\rangle$ of the Laplacian on $H^2$, as written in (\ref{eq:upm}), by letting $\eta = i \theta$ and taking $l \to i p - 1/2$ for an arbitrary real number $p$.  The analytic continuation is
\begin{equation}
	|Y_l^m\rangle \to c_{pm}|u_{pm}\rangle~,
\label{eq:yumap}
\end{equation}
with some overall constants $c_{pm}$. The regularity condition at the poles $\theta=0,\pi$ is the reason eigenvalues are discrete on $S^2$ and on the non-compact hyperbolic plane $H^2$ their is no analogous condition. 

To summarize, the map (\ref{eq:yxmap}) identifies the dilaton profiles on $S^2$ and $H^2$ by analytic continuation.  These profiles in turn generate the $SU(2) \times SU(2)$ conformal Killing algebra on $S^2$ (\ref{eq:su2su2}) and the $SL(2,\mathbb{R}) \times SL(2,\mathbb{R})$ conformal Killing algebra on $H^2$ (\ref{eq:sl2sl2}). Explicit comparison of the respective generators map them onto one another as:
\begin{equation}
	\begin{split}\begin{aligned}[t]
		L_z^{(S^2)} &\to -i L_0^{(H^2)}~, \quad \\
		R_z^{(S^2)} &\to -i R_0^{(H^2)}~, \quad
	\end{aligned}
	\begin{aligned}[t]
		L_\pm^{(S^2)} &\to - L_{\pm 1}^{(H^2)}~, \\
		R_\pm^{(S^2)} &\to - R_{\pm 1}^{(H^2)}~.
	\end{aligned}\end{split}
\end{equation}
In other words, the well-known enhancement of symmetry for the Hydrogen atom analytically continues into the emergence of conformal isometry that we found in subsection~\ref{sec:jt:euclid} for the Jackiw-Teitelboim model.  

As an application, we now use analytic continuation as a tool to compute dilaton matrix elements on $H^2$ via analytic continuation from $S^2$.
In particular, we want to prove the claim (\ref{eq:r0vanish}) that the matrix element $\langle u_{pm}| R_0 |u_{pm} \rangle = 0$ on $H^2$. This amounts to
$\langle Y_l^m | R_z | Y_{l}^{m}\rangle$ on $S^2$, up to overall constants.  Our explicit expression for $R_z$ gives
\begin{equation}\begin{aligned}
\langle Y_l^m | R_z | Y_{l'}^{m'}\rangle &= \int d^2x\,\sqrt{g}\, Y_l^{m *} \left(-\sin\theta\,\partial_\theta\right) Y_{l'}^{m'} \\
	 &= (l+2)\sqrt{\frac{(l-m+1)(l+m+1)}{(2l+1)(2l+3)}} \delta_{l,l'-1} \delta_{m,m'} \\
	 &\quad - (l-1)\sqrt{\frac{(l-m)(l+m)}{(2l-1)(2l+1)}} \delta_{l,l'+1}\delta_{m,m'}~,
\end{aligned}\end{equation}
where we used the Legendre polynomial identity
\begin{equation}\begin{aligned}
	\int_{-1}^{+1}dx\,(1-x^2) P_l^m(x) \partial_x P_{l'}^{m}(x) &= \frac{2(l+2)(l+m+1)!}{(2l+1)(2l+3)(l-m)!} \delta_{l,l'-1} \\
	&\quad - \frac{2(l-1)(l+m)!}{(2l-1)(2l+1)(l-m-1)!} \delta_{l,l'+1}~.
\end{aligned}\end{equation}
In particular, we find that the matrix element is non-zero only when $l$ and $l'$ are not equal, and so $\langle Y_l^m | R_z | Y_l^m\rangle = 0$.  The analytic continuation back to $H^2$ then immediately gives
\begin{equation}
	\langle u_{pm} | R_0 | u_{pm}\rangle = 0~,
\end{equation}
confirming that the dilaton matrix elements that show up in the one-loop continuous mode partition function of the Jackiw-Teitelboim model vanish. 

The interpretation of this result, in analogy with the Hydrogen atom, is that within a level specified by a given principal quantum number $n$, 
the angular momentum operator ${\bf L}$ relates states with the same orbital quantum number $l$, while the ``hidden" symmetry operator ${\bf R}$ relates those with
different values of $l$.

\bibliographystyle{JHEP}
\bibliography{JTLogsFinal}

\end{document}